\documentclass[12pt]{article}
\usepackage{amsthm,amsmath,amssymb}
\usepackage{multirow}
\usepackage[pdftex]{graphicx}
\usepackage{subfigure}
\usepackage{makecell}
\usepackage{booktabs}
\usepackage{array}
\usepackage{fullpage}
\usepackage{url}
\usepackage{algorithm}
\usepackage{algorithmic}
\usepackage{bm}
\usepackage{smile}
\usepackage{mathtools}
\usepackage{wrapfig}
\usepackage{mathrsfs}
\usepackage{color}
\usepackage{verbatim}
\usepackage{enumerate}
\usepackage{bbm}
\usepackage{setspace}
\usepackage{multirow}
\usepackage{nameref}

\usepackage[usenames,dvipsnames,svgnames,table]{xcolor}
\usepackage[colorlinks=true,
            linkcolor=blue,
            urlcolor=blue,
            citecolor=blue]{hyperref}

\usepackage[plain]{fancyref}
\makeatletter
\def\mkfancyprefix#1#2{%
\expandafter\def\csname fancyref#1labelprefix\endcsname{#1}%
\begingroup\def\x{\endgroup\frefformat{plain}}%
    \expandafter\x\csname fancyref#1labelprefix\endcsname
    {\MakeLowercase{#2}\fancyrefdefaultspacing##1}%
\begingroup\def\x{\endgroup\Frefformat{plain}}%
    \expandafter\x\csname fancyref#1labelprefix\endcsname
    {#2\fancyrefdefaultspacing##1}%
\begingroup\def\x{\endgroup\frefformat{vario}}%t
    \expandafter\x\csname fancyref#1labelprefix\endcsname
    {\MakeLowercase{#2}\fancyrefdefaultspacing##1##3}%
\begingroup\def\x{\endgroup\Frefformat{vario}}%
    \expandafter\x\csname fancyref#1labelprefix\endcsname
    {#2\fancyrefdefaultspacing##1##3}%
}
\makeatother

\mkfancyprefix{thm}{Theorem}
\mkfancyprefix{lem}{Lemma}
\mkfancyprefix{prop}{Proposition}
\mkfancyprefix{cor}{Corollary}
\mkfancyprefix{def}{Definition}
\mkfancyprefix{cl}{Claim}
\mkfancyprefix{ss}{Subsection}
\mkfancyprefix{alg}{Algorithm}
\mkfancyprefix{ass}{Assumption}

\newcommand{\blind}{0}

\begin{document}

\def\spacingset#1{\renewcommand{\baselinestretch}%
{#1}\small\normalsize} \spacingset{1}

\newcounter{mylabelcounter}

\makeatletter
\newcommand{\labelText}[2]{%
#1\refstepcounter{mylabelcounter}%
\immediate\write\@auxout{%
  \string\newlabel{#2}{{1}{\thepage}{{\unexpanded{#1}}}{mylabelcounter.\number\value{mylabelcounter}}{}}%
}%
}
\makeatother

\numberwithin{equation}{section}
\numberwithin{theorem}{section}
\numberwithin{assumption}{section}

%%%%%%%%%%%%%%%%%%%%%%%%%%%%%%%%%%%%%%%%%%%%%%%%%%%%%%%%%%%%%%%%%%%%%%%%%%%%%%

  \title{\bf Covariance-based sample selection for heterogeneous data: 
Applications to gene expression and autism risk gene detection}
  \author{Kevin Z. Lin\\
  {\footnotesize Carnegie Mellon University, Department of Statistics \& Data Science, Pittsburgh, PA} \\
    Han Liu  \\
    {\footnotesize Northwestern University, Department of Electrical Engineering and Computer Science, Evanston, IL}
    \\
    Kathryn Roeder \\
    {\footnotesize Carnegie Mellon University, Department of Statistics \& Data Science, Pittsburgh, PA}\thanks{Han Liu's research is supported by the NSF BIGDATA 1840866, NSF CAREER 1841569, NSF TRIPODS 1740735, DARPA-PA-18-02-09-QED-RML-FP-003, along with an Alfred P Sloan Fellowship and a PECASE award. Kathryn Roeder's research is
    supported by NIMH grants R37MH057881 and U01MH111658-01.}\\
        }
  \maketitle

%\bigskip
\begin{abstract}
Risk for autism can be influenced by genetic mutations in hundreds of genes.  
Based on findings showing that genes with highly correlated gene expressions
are functionally interrelated, ``guilt by association'' methods such as DAWN have been developed to identify these autism risk genes.  
Previous research analyzes the BrainSpan dataset, which contains gene expression of brain tissues from varying regions and developmental periods. 
Since the spatiotemporal properties of brain tissue is known to affect the gene expression's covariance, previous research have focused only on a specific subset of samples to avoid the issue of heterogeneity.
This leads to a potential loss of power when detecting risk genes. 
In this article, we develop a new method called COBS (COvariance-Based sample Selection) to find 
a larger and more homogeneous subset of samples that share the same population covariance matrix for the downstream DAWN analysis. 
To demonstrate COBS’s effectiveness, we utilize genetic risk scores from two sequential data
freezes obtained in 2014 and 2019. We show COBS improves DAWN’s ability to predict risk genes detected in the newer data freeze when utilizing the risk scores of
the older data freeze as input.
\end{abstract}

\noindent%
\vfill

\newpage
%\spacingset{1.45} % DON'T change the spacing!

\section{Introduction} \label{sec:intro}

The genetic cause of autism spectrum disorder (ASD), a neurodevelopmental disorder that affects roughly 1-2\% 
individuals in the United States, remains an open problem despite decades of 
research \citep{autism2014prevalence}. 
ASD is characterized primarily by
impaired social functions and repetitive behavior \citep{kanner1943autistic, rutter1978diagnosis}. 
To better understand this disorder,
scientists identify specific genes that are liable for increasing the chance of developing ASD when damaged or mutated \citep{sanders2015insights}.
These are genes are called risk genes.
While breakthroughs in genomic technologies and the
availability of large ASD cohorts have led to the discovery of dozens of risk genes,
preliminary studies suggest there are hundreds of risk genes still unidentified 
\citep{buxbaum2012autism}.  
In this work, we build upon the current statistical methodologies to further improve our ability to
identify risk genes.

We focus on statistical methods that use gene co-expression networks to help identify risk genes.
These networks are estimated from  brain tissue's gene expression data.
Since these gene co-expression networks provide insight into genes that regulate normal biological mechanisms in fetal and early brain development, it was hypothesized
that risk genes that alter these mechanisms 
should be clustered in these networks \citep{vsestan2012emerging}. 
Early findings confirmed this hypothesis 
\citep{parikshak2013integrative, willsey2013coexpression}. 
These results led to the development of the Detection Association With Networks (DAWN) algorithm which 
uses a ``guilt by association'' strategy -- implicating new risk genes based on their connectivity to previously identified risk genes
\citep{liu2014dawn, liu2014network}. 
However, 
the previous DAWN analyses suffer from statistical limitations that we will investigate
and resolve in this article.

We challenge previous analyses' assumptions regarding 
the homogeneity of the covariance matrix in gene expression data.
Previous DAWN analyses assume that
gene expression samples from the same brain tissue type
share the same covariance matrix. 
This assumption was influenced by the findings in \cite{kang2011spatio} and \cite{willsey2013coexpression},
which showed that
 gene co-expression patterns differ among different brain regions
and developmental periods on average. Statistically, this means that the 
covariance matrix among the gene expressions 
may differ with respect to the spatiotemporal properties of the brain tissue.
Hence,  previous DAWN analyses \citep{liu2014dawn, liu2014network} use only samples from a particular brain tissue type chosen
by the findings in \cite{willsey2013coexpression}.
However,
no further statistical analysis is performed to check for homogeneity
of this specific subset of samples.
In addition, since previous DAWN analyses limit themselves to a subset of
gene expression samples, 
many other samples assumed to be heterogeneous are excluded.
This leads to a potential loss of power when identifying risk genes.  

To overcome these limitations, we develop a method called COBS (COvariance-Based sample Selection),
a two-staged procedure in order to select a subset of gene expression samples 
in a data-driven way 
that is more homogeneous and larger in sample size than the fixed subset used
previously. In the first stage, 
we take advantage of the recent developments in high-dimensional
covariance testing \citep{cai2013two,chang2015comparing} to 
determine whether if the gene expression from two different brain tissues share the same population
covariance matrix. We combine this with a multiple-testing method called Stepdown that 
accounts for the dependencies among many hypothesis tests \citep{romano2005exact, chernozhukov2013gaussian}.
In the second stage, after determining which pairs of brain tissues have statistically indistinguishable
covariance matrices, we develop a clique-based procedure to select which brain tissues to use in the downstream
DAWN analysis.
We show that COBS selects brain tissues within the BrainSpan dataset that align with
current scientific knowledge and also
leads to an improved gene network estimate for implicating risk genes.
This article addresses the numerous algorithmic challenges needed to implement this idea.

In Section 2, we describe the data and statistical model for heterogeneity in the covariance matrix. 
In Section 3, we provide a visual diagnostic to investigate the homogeneity assumptions of previous DAWN analyses.
In Section 4, we describe the different stages of COBS to find 
a subset of homogeneous samples within a dataset.
In Section 5, we illustrate the properties of COBS on synthetic datasets.
In Section 6, we apply our procedure on gene expression data to show that,
when combined with DAWN, we have an improved gene network that can better
implicate risk genes.
Section 7 provides an overall summary and discussion.

\section{Data and model background} \label{sec:model}

Due 
to the challenge of obtaining and preserving brain tissue,
datasets recording the gene expression patterns of brain tissue are rare. 
The BrainSpan project 
contributes one of the largest 
microarray expression datasets available (the ``BrainSpan dataset'' henceforth), 
sampling tissues from 57 postmortem brains with no signs of large-scale genomic
abnormalities  \citep{kang2011spatio}. 
Many studies have favored this dataset because its 1294 microarray samples
capture the spatial and temporal changes in gene expression that occur in the brain
during the entirety of development \citep{de2014synaptic,dong2014novo,cotney2015autism}.
While our paper focuses on this particular microarray expression dataset, our method
would apply to other gene expression datasets such as RNA sequencing data.

\begin{figure}
  \centering
 \includegraphics[width=450px]{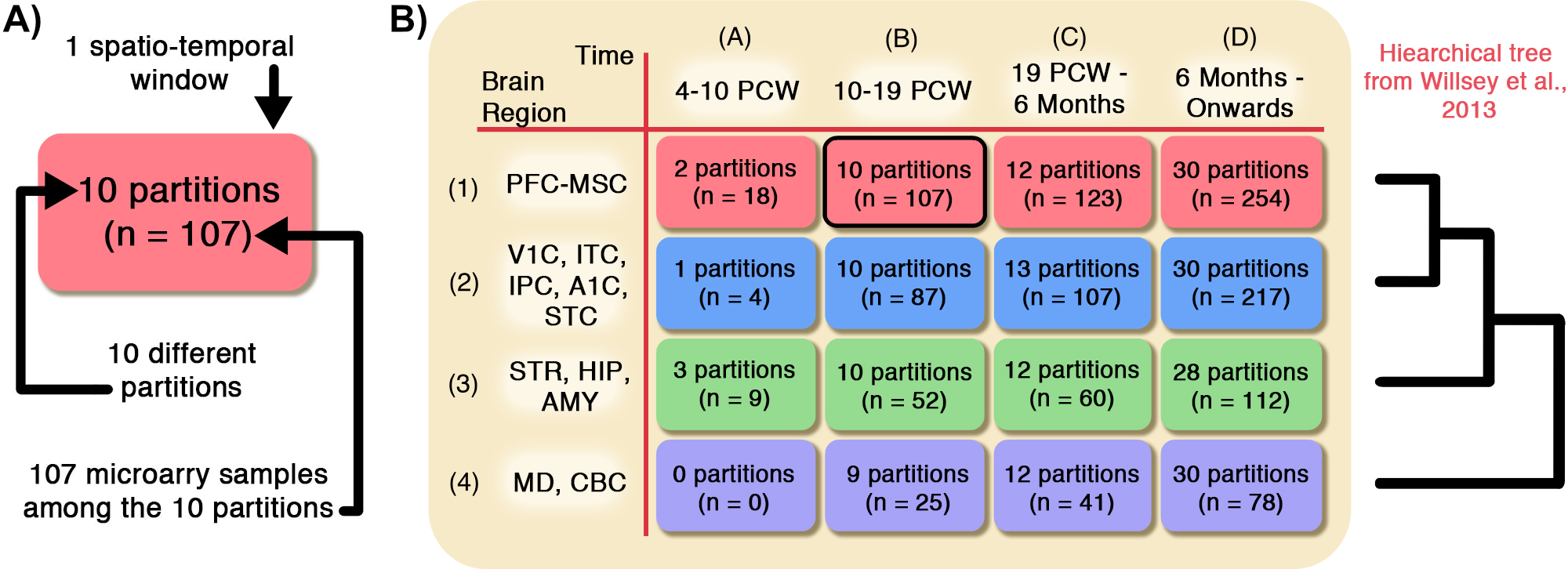}
  \caption
   {\small (A)
   107 microarray
   samples grouped by the originating 10 brains. This forms 10 different
   partitions. Since all these partitions originate from the same brain
   region and developmental period, they are further grouped into the 
   same window. (B) 
   The 57 postmortem brains belong to 4 different developmental periods (columns). Here, PCW stands for
    post-conceptual weeks.
  Each brain is dissected and sampled at 4 different brain regions (rows).
  In total, over the 212 partitions, there are 1294 microarray samples, each measuring the expression
of over 13,939 genes.
   Window 1B (outlined in black) is the window that previous work 
   \citep{liu2014network} focus on, and the
hierarchical tree from \cite{willsey2013coexpression} is shown to the right.
Additional details about the abbreviations are given in Appendix \ref{sec:appdata}.}
    \label{fig:partpreselect}
\end{figure}

The heterogeneity of gene expression due to the spatiotemporal differences 
in brain tissues presents statistical challenges.
As documented in \cite{kang2011spatio}, the region and developmental period
 of the originating brain tissue
 contribute more to the heterogeneity than other variables such as sex and ethnicity.
To understand this heterogeneity, we use the following schema to model the BrainSpan dataset.
Each of the 1294 microarray samples is categorized into one of 16 
\emph{spatiotemporal windows}, or \emph{windows} for short, depending on
which brain region and developmental period the brain tissue is derived from.
Within each window, all microarray samples originating from the same brain
are further categorized into the same 
\emph{partition}. 
There are 212 partitions in total.
\Fref{fig:partpreselect} summarizes how many partitions and microarray samples belong in
each window in the BrainSpan dataset.
This schema allows us to model the microarray samples more realistically since 
the gene co-expression patterns vary greatly on average from window to window \citep{willsey2013coexpression}.
Additionally, \cite{willsey2013coexpression} find that among all the windows, the known
risk genes in Window 1B are most tightly co-expressed.
Window 1B is highlighted in \Fref{fig:partpreselect} and contains
the 107 microarray samples
from the prefrontal cortex and primary motor-somatosensory cortex
from 10 to 19 post-conceptual weeks.
Due to this finding, previous DAWN analyses focus on 
all 107 samples from 10 partitions, assuming
that these samples were homogeneous without further statistical investigation, and discard the remaining 1187 samples, \citep{liu2014dawn, liu2014network}. We seek to improve upon this heuristical
sample selection procedure, first by formalizing a statistical model.

\subsection{Statistical model}
We use a mixture model 
that assumes that microarry samples from the same partition are homogeneous
while samples from different partitions could be heterogeneous. 
For the $p$th partition,
let \smash{$\bX^{(p)}_1, \ldots \bX^{(p)}_{n_p} \in \mathbb{R}^d$} denote $n_p$ 
i.i.d. samples, and let $w(p)$ denote the window that partition $p$ resides in. 
These $n_p$ samples
are drawn from either a distribution with covariance
$\bSigma$,
or another distribution with a different covariance matrix $\bSigma_p$.
Our notation emphasizes that the distributions in consideration are not necessarily Gaussian, and
$\bSigma$ is the covariance matrix shared among
all partitions, while $\bSigma_p$ may vary from partition to partition.
A fixed but unknown parameter \smash{$\gamma_{w(p)} \in [0,1]$} 
controls how frequently the partitions in window $w$ are drawn from
these two distributions, meaning
it controls the amount of heterogeneity.
For each partition $p$, this mixture model is succinctly described as, 
\begin{align}
I^{(p)} &\sim \text{Bernoulli}(\gamma_{w(p)}), \nonumber\\
\bX^{(p)}_1, \ldots, \bX^{(p)}_{n_{p}}&\stackrel{i.i.d.}{\sim}
 \begin{cases}
    D(\bSigma)      & \quad \text{if } I^{(p)} = 1\\
   D(\bSigma_{p})  & \quad \text{otherwise},\\
  \end{cases} \label{eq:hiermodel}
\end{align}
where $D(\bSigma)$ denotes an arbitrary distribution with covariance matrix $\bSigma$, and $I^{(p)}$ is the latent variable that determines whether or not the samples in
partition $p$ have covariance $\bSigma$ or $\bSigma_p$.
With this model setup, our task is to determine the set of partitions that originate from the covariance matrix $\bSigma$, which we will call
\begin{equation} \label{eq:p}
\mathcal{P}=\Big\{p ~:~ I^{(p)} = 1\Big\}.
\end{equation}

The findings of \cite{kang2011spatio} and \cite{willsey2013coexpression} inform us on how
much heterogeneity to expect within a window via $\gamma_{w(p)}$.
While analyses such as \cite{liu2014network} assume that 
all the samples in Window 1B are homogeneous,
it is noted in \cite{kang2011spatio} that 
sampling variability in brain dissection and
in the proportion of white and gray matter in different brain tissues
can cause variability in the gene co-expression patterns.
This means that scientifically, we do not expect all the partitions in Window 1B to
be homogeneous (i.e., $\gamma_{w(p)} = 1$).
Furthermore, \cite{willsey2013coexpression} 
find a hierarchical clustering among the four brain regions.
This is illustrated in \Fref{fig:partpreselect}, where the gene 
co-expression patterns in the brain regions represented in
first row are most similar to those in the second row and least similar to
those in the fourth row.
The authors also find a smooth continuum of gene expression patterns across 
different developmental periods, represented as the columns of the table in \Fref{fig:partpreselect}.
Hence, we
expect $\gamma_{w(p)}$ to decrease smoothly as the window $w$ becomes more dissimilar to Window 1B, in both the spatial and temporal direction. 

\subsection{Connections to other work}

Other work use models similar to \eqref{eq:hiermodel} on microarray expression data
to tackle the different co-expression patterns among different tissues and subjects, but their methods
differ from ours. 
One direction is to directly cluster the covariance matrices of each partition \citep{ieva2016covariance}.
However, this approach does not account for the variability in the empirical covariance matrix,
unlike our hypothesis-testing based method.
Another approach is to explicitly model the population covariance matrix for each partition 
as the summation of a shared
component and a partition-specific heterogeneous component.
This is commonly used in batch-correction procedures where 
the analysis removes the heterogeneous component from each partition \citep{leek2007capturing}. 
However, we feel such an additive model is too restrictive for analyzing the BrainSpan dataset, as we 
do not believe there is a shared covariance matrix across all windows of the brain. 
Instead, our approach will find specific set of partitions with statistically indistinguishable covariance
matrices.

\section{Elementary analysis} \label{sec:elementary}

In this section, we develop a visual diagnostic to investigate if the 10 partitions in Window 1B used in previous work \citep{liu2014dawn, liu2014network} are
as homogeneous as these previous analyses assume.
Using a hypothesis test for equal covariances, our diagnostic leverages the following idea:
we divide the partitions into two groups and apply a hypothesis test to the samples 
between both groups.
If all the partitions were truly drawn from distributions with equal covariances, then 
over many possible divisions, the empirical distribution of the resulting p-values should be roughly uniform.
We can visualize this distribution by using a QQ-plot.
The less uniform the p-values look, the less we are inclined to interpret our partitions to be all
drawn from distributions with equal covariances. The following algorithm summarizes this diagnostic.

\vspace{1em}
\textbf{\labelText{Algorithm 1}{label:alg1}: Covariance homogeneity diagnostic}
\vspace{-.5em}
\begin{enumerate}
\item Loop over trials $t = 1, 2, \ldots, T$:
\begin{enumerate}
\item Randomly divide the selected partitions in the set $\hat{\mathcal{P}}$ into two sets,
$\hat{\mathcal{P}}^{(1)}$ and $\hat{\mathcal{P}}^{(2)}$, such that 
$\hat{\mathcal{P}}^{(1)} \cup \hat{\mathcal{P}}^{(2)} = \hat{\mathcal{P}}$ and $\hat{\mathcal{P}}^{(1)} \cap \hat{\mathcal{P}}^{(2)} = \emptyset$.
\item For each partition $p \in \hat{\mathcal{P}}^{(1)}$, center the samples $\bX^{(p)}_1, \ldots, 
\bX^{(p)}_{n_p}$. Then aggregate all samples in $\hat{\mathcal{P}}^{(1)}$ 
to form the set of samples
\[
\mathcal{X}= \bigcup_{p \in \hat{\mathcal{P}}^{(1)}} \big\{\bX^{(p)}_1, \ldots, \bX_{n_p}^{(p)} \big\}.
\]
Similarly, form the set of samples $\mathcal{Y}$ from the set of partitions $\hat{\mathcal{P}}^{(2)}$.

\item Compute the p-value for a hypothesis test that tests whether or not the
samples in $\mathcal{X}$ and $\mathcal{Y}$
have the same covariance matrix.
\end{enumerate}
\item Produce a QQ-plot of the resulting $T$ p-values to see if empirical distribution of the p-values is
close to a uniform distribution.
\end{enumerate}

We remind the reader that the above procedure is a diagnostic.
This is not necessarily a recipe for a goodness-of-fit test since the $T$ p-values are not independent,
which makes it difficult to analyze its theoretical properties without a carefully designed global null test.
However, as we will demonstrate in later sections of this article, this diagnostic is nonetheless
able to display large-scale patterns in our dataset.

\subsection{Specification of covariance hypothesis test} \label{ss:specification}
To complete the above diagnostic's description, we describe the procedure to test for equality of covariance matrices. Following the model \eqref{eq:hiermodel}, let
$\mathcal{X} = \{\bX_1, \ldots, \bX_{n_1}\}$ and $\mathcal{Y} = \{\bY_1, \ldots, \bY_{n_2}\}$ 
be $n_1$ and 
$n_2$ i.i.d. samples from $d$-dimensional distribution with covariance $\bSigma_{X}$ and $\bSigma_{Y}$ respectively, both with an empirical mean of $\mathbf{0}$.
We define $\mathbb{X} \in \mathbb{R}^{n_1 \times d}$ and $\mathbb{Y} \in \mathbb{R}^{n_2\times d}$
as the matrices formed by concatenating these samples row-wise.
Define the empirical covariance matrices as \smash{$\hat{\bSigma}_X = \mathbb{X}^\top\mathbb{X}/n_1$},
and \smash{$\hat{\bSigma}_Y =\mathbb{Y}^\top\mathbb{Y}/n_2$},
where we denote the individual elements of these matrices as 
\smash{$\hat{\bSigma}_X = [\hat{\sigma}_{X,ij}]_{1\leq
i,j\leq d}$} and likewise for $\hat{\bSigma}_Y$. 

We now discuss the hypothesis test for equal covariance, $H_0: \bSigma_{X} = \bSigma_Y$,
that we will consider in this article based on the test statistic defined in \cite{chang2015comparing}
which extends \cite{cai2013two}.
In these works, the authors note that if $\bSigma_X = \bSigma_Y$, then
the maximum element-wise difference between $\bSigma_X$ and $\bSigma_Y$ is 0.
Hence, \cite{chang2015comparing} defines the test statistic $\hat{T}$ as the maximum of
squared element-wise differences between $\hat{\bSigma}_X$
and $\hat{\bSigma}_Y$, normalized by its variance.
Specifically, 
\begin{equation} \label{eq:test_stat}
\hat{T} = \max_{ij}\big(\hat{t}_{ij}\big) \quad \text{where
}\hat{t}_{ij}=
\frac{\big(\hat{\sigma}_{X,ij}-\hat{\sigma}_{Y,ij}\big)^2}{\hat{s}_{X,ij}/n_1 +
\hat{s}_{Y,ij}/n_2}, \quad i,j \in 1,\ldots, d,
\end{equation}
where 
\smash{$\hat{s}_{X,ij} =
\sum_{m=1}^{n_1}(\mathbb{X}_{mi}\mathbb{X}_{mj}-\hat{\sigma}_{X,ij})^2/n_1$} 
is the empirical
variance of the variance-estimator $\hat{\sigma}_{X,ij}$, and $\hat{s}_{Y,ij}$
is defined similarly. 

Then, \cite{chang2015comparing} constructs an empirical null distribution of $\hat{T}$ 
under $H_0: \bSigma_{X} = \bSigma_Y$
using the multiplier bootstrap \citep{chernozhukov2013gaussian}.
On each of the $b \in \{1, \ldots, B\}$ trials, the multiplier bootstrap 
computes a bootstrapped test statistic $\hat{T}^{(b)}$ by 
weighting each of the $n_1+n_2$ observations by a standard Gaussian random
variable drawn independently of all other variables, denoted collectively as
\smash{$(g^{(b)}_1,\ldots,g_{n_1}^{(b)},g_{n_1+1}^{(b)},\ldots,g_{n_1+n_2}^{(b)})$}.
Specifically, we construct the bootstrap statistic for the $b$th trial as
\begin{equation} \label{eq:bootstrap_stat}
\hat{T}^{(b)} = \max_{ij}\big(\hat{t}^{(b)}_{ij}\big) \quad \text{where }\hat{t}^{(b)}_{ij}=
\frac{\big(\hat{\sigma}^{(b)}_{X,ij}-\hat{\sigma}^{(b)}_{Y,ij}\big)^2}{\hat{s}_{X,ij}/n_1
+ \hat{s}_{Y,ij}/n_2}, \quad i,j \in 1,\ldots,d,
\end{equation}
where \smash{$\hat{\sigma}^{(b)}_{X,ij} =
\sum_{m=1}^{n_1}g_m^{(b)}(\mathbb{X}_{mi}\mathbb{X}_{mj}
- \hat{\sigma}_{X,ij})/n_1$} and \smash{$\hat{\sigma}^{(b)}_{Y,ij} =
\sum_{m=1}^{n_2}g_{n_1+m}^{(b)}(\mathbb{Y}_{mi}\mathbb{Y}_{mj}-\hat{\sigma}_{Y,ij})/n_2$}. 
We compute the p-value by counting the proportion of bootstrap statistics that are larger than the test statistic, 
\[
\text{p-value} = \frac{\big|\{b : |\hat{T}^{(b)}| \geq |\hat{T}|\}\big|}{B}.
\]
\cite{chang2015comparing} prove that this test
has asymptotically $1-\alpha$ coverage under the null hypothesis
as long as the all distributions in the
distribution family $D$ in \eqref{eq:hiermodel} have sub-Gaussian and sub-exponential tails,
even in the high-dimensional
regime where $d \gg \max(n_1, n_2)$.

\subsection{Application to BrainSpan}
\begin{figure}[tb]
  \centering
  \includegraphics[width=400px]{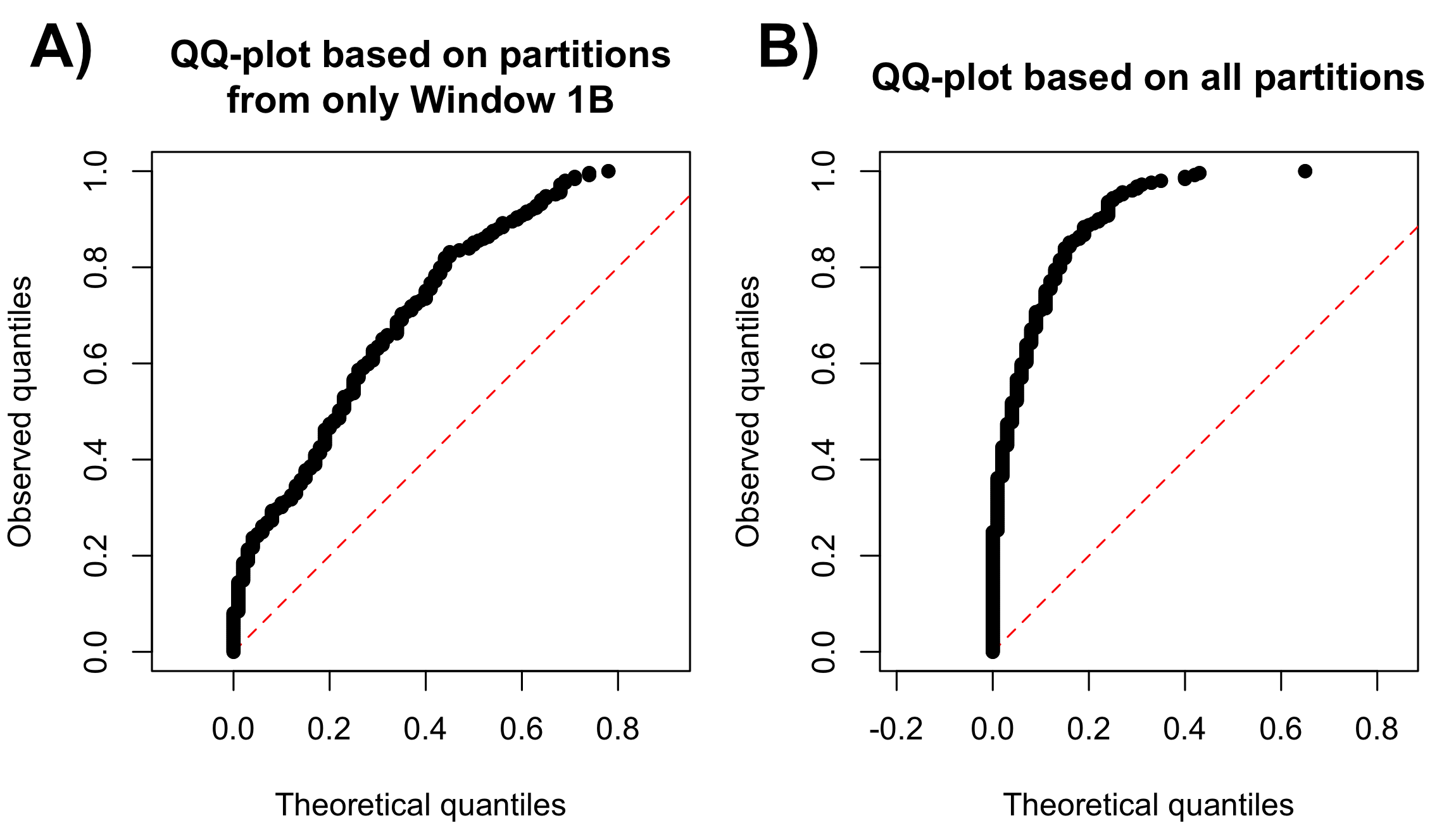}
  \vspace{-1em}
  \caption
   {\small QQ-plots of the 250 p-values generated when applying our diagnostic to the BrainSpan dataset.
   (A) The diagnostic using only the partitions in Window 1B, showing a moderate amount of
   heterogeneity. (B) The diagnostic  using all 125 partitions 
   in the BrainSpan dataset, showing a larger amount of heterogeneity.}
    \label{fig:initial_qq}
\end{figure}

Equipped with a complete description of the diagnostic, we apply it to the BrainSpan dataset.
Among the 10 partitions in Window 1B, we divide the partitions into 
two groups uniformly at random 250 times, and compute a p-value using
Method 1 (with normalization) for each division using 200 bootstrap trials. The QQ-plot of the resulting
p-values are shown in \Fref{fig:initial_qq}A, where we see that the p-values are biased towards 0.
This implies the 10 partitions in Window 1B are heterogeneous since
they do not seem to all share the same covariance matrix.
Furthermore, we apply this diagnostic to all partitions in the BrainSpan dataset with 5 or more samples.
This results in using only 125 of the 212 partitions shown in \Fref{fig:partpreselect}.
The resulting p-values become more biased towards 0 (\Fref{fig:initial_qq}B), implying
the dataset as a whole is more heterogeneous than the partitions in Window 1B.
In the next section, we
develop a method to resolve this issue by finding the largest subset of partitions 
possible among the 125 partitions in the BrainSpan dataset that
share the same covariance matrix.

\section{Methods: COBS (Covariance-based sample selection)} \label{sec:method}
While we have discussed a method to test for
equivalent covariance matrices between any two partitions in \Fref{sec:elementary}, 
we cannot directly apply
this method to select a large number of homogeneous partitions in the BrainSpan dataset without suffering a loss of power
due to multiple testing.
Since there are $r = 125$ partitions with more than 5 samples, applying the hypothesis test to each pair of
partitions results in ${r \choose 2} = 7750$ dependent p-values.
These p-values are dependent since each of the $r$ partitions is involved in $r-1$ hypothesis tests.
Hence, standard techniques such as 
a Bonferroni correction are too conservative when accounting for these dependencies, likely
leading to a loss of power.

To properly account for this dependency, we introduce our new method called COBS, which comprises of two
parts. 
First, we use a Stepdown method in \Fref{ss:test}
that simultaneously tests all ${r \choose 2}$ hypothesis tests for
equal covariance matrices, which builds upon the bootstrap test 
introduced previously in \Fref{sec:elementary}.
After determining which of the ${r \choose 2}$ pairs of partitions do not
have statistically significant differences in their covariance matrices, we develop
a clique-based method in \Fref{ss:selection} to select a specific set of partitions $\hat{\mathcal{P}}$.

\subsection{Stepdown method: multiple testing with dependence} \label{ss:test}

We use a Stepdown method developed in
\cite{chernozhukov2013gaussian} to control the family-wise error rate (FWER). 
We tailor the bootstrap-based test in \Fref{ss:specification} to our specific setting in the algorithm below.
We denote $\hat{T}_{(i,j)}$ as the test statistic formed using \eqref{eq:test_stat} to test if the covariance of samples between partition $i$ 
and partition $j$ are equal. Similarly, let $\hat{T}^{(b)}_{(i,j)}$ denote the corresponding bootstrap statistics on the $b$th bootstrap trial. Here, $\text{quantile}(\{x_1, \ldots, x_n\}; 1-\alpha)$ represents the empirical $(1-\alpha)\cdot 100$\% quantile
of the vector $(x_1, \ldots, x_n)$.

\vspace{1em}
\textbf{\labelText{Algorithm 2}{label:alg2}: Stepdown method}
\begin{enumerate}
\item Initialize the list enumerating all \smash{${r \choose 2}$} null hypotheses corresponding to the 
set of partition pairs, $\mathcal{L} = \Big\{(1,2), \ldots, (r-1,r)\Big\}$. 
\item Calculate $\hat{T}_\ell$ for each $\ell \in \mathcal{L}$, as stated in \eqref{eq:test_stat}.
\item Loop over steps $t = 1, 2, \ldots$:
\begin{enumerate}
\item For each bootstrap trial $b = 1,\ldots,B$:
\begin{enumerate}
\item Generate $N = \sum_p n_p$ i.i.d. standard Gaussian random variables, one for each sample in each partition, and compute $\hat{T}^{(b)}_\ell$ for all $\ell \in \mathcal{L}$, as stated in \eqref{eq:bootstrap_stat}.
\item Compute 
\begin{equation} \label{eq:max_bootstrap}
\hat{T}^{(b)} = \max\Big\{\hat{T}^{(b)}_\ell: \ell \in \mathcal{L}\Big\}.
\end{equation}
\end{enumerate}
\item Remove any $\ell \in \mathcal{L}$ if
\[
\hat{T}_\ell \geq \text{quantile}\Big(\{\hat{T}^{(1)}, \ldots, \hat{T}^{(b)}\}; 1-\alpha \Big). 
\]
If no elements are removed from $\mathcal{L}$, return the null hypotheses corresponding to $\mathcal{L}$. Otherwise, continue to step $t+1$.
\end{enumerate}
\end{enumerate}

Using techniques in \cite{romano2005exact} and \cite{chernozhukov2013gaussian}, it can be proven that 
this method has the following asymptotic FWER guarantee,
\begin{equation} \label{eq:stepdown}
\mathbb{P}\Big(\text{no true null hypothesis among }\mathcal{H}
\text{ null hypotheses are rejected}\Big) \geq 1 - \alpha + o(1)
\end{equation}
under the same assumptions posed in \cite{chang2015comparing}.
The reason \nameref{label:alg2} is able to control the FWER without a Bonferroni 
correction is because the null distribution in the Stepdown method is properly calibrated to account for the 
joint dependence among the ${r \choose 2}$ tests.
Specifically, when ${r \choose 2}$ tests are individually performed as in \Fref{ss:specification}, the
test statistics \eqref{eq:test_stat} are dependent, but the 
bootstrapped null distributions do not account for this dependence.
Hence, accounting for the dependence via a Bonferroni correction after-the-fact can lead to a substantial loss in power.
However, in the Stepdown procedure, the bootstrapped null distributions retain the dependencies jointly
since they are generated from the same $N$ Gaussian random variables in each trial.
See \cite{chernozhukov2013gaussian} (Comment 5.2) for a further discussion.

\paragraph{Robustness concerns.} 
In practice, due to the maximum function in the test statistic $\hat{T}_\ell$ displayed in \eqref{eq:test_stat}, the Stepdown method could
possibly erroneously reject a hypothesis due to the presence of outliers.
One way to circumvent this problem to purposely shrink the value of the test statistic
 $\hat{T}_\ell$ while leaving the bootstrapped statistics $\hat{T}^{(b)}_\ell$ in \eqref{eq:bootstrap_stat} the same. Specifically, we can replace
 $\max_{ij}(\hat{t}_{ij})$ in \eqref{eq:test_stat} with the $\text{quantile}(\{\hat{t}_{ij}\}_{ij}; 1-\epsilon)$, where $\epsilon$ is a positive number extremely close to 0. This has the desired effect of ``discarding'' the large values in $\{\hat{t}_{ij}\}_{ij}$. 
Observe that this procedure would potentially lead to a slight loss in power, but the inferential guarantee in \eqref{eq:stepdown} still holds since there can only be strictly less rejections.
 
\paragraph{Computational concerns.} 
While we use
the test statistics \eqref{eq:test_stat} when describing the Stepdown method,
we note that this method applies to a broader family of test statistics. 
In Appendix \ref{sec:accelerated}, we discuss in detail one alternative to the test statistic in  \eqref{eq:test_stat} that
can dramatically reduce up the computation complexity of the Stepdown method. However, we defer this to
the appendix because in our specific problem setting of testing equality of covariances, it does not seem to
perform well empirically.

\subsection{Largest quasi-clique: selecting partitions based on testing results} \label{ss:selection}

After applying the covariance testing with the Stepdown method described in the previous subsection, we have a subset of null hypotheses from $\mathcal{H}$ that we accepted. In this subsection, we develop a clique-based method to estimate $\mathcal{P}$, the subset of partitions that share the same covariance matrix defined in \eqref{eq:p}, from our accepted null hypotheses.

We conceptualize the task of selecting partitions as selecting vertices from a graph that form a dense subgraph. 
Let $H_{0,(i,j)}$ denote 
the null hypothesis that the population covariance matrices for partition $i$ and $j$ are equal. 
Let $G = (V,E)$ be an undirected graph with vertices $V$ and edge set $E$ such that
\begin{equation}\label{eq:hypothesis_graph}
V = \big\{1,\ldots,r\big\}, \quad E = \big\{(i,j) : H_{0,(i,j)} \text{ is accepted by the Stepdown method}\big\}.
\end{equation}
Since each of the \smash{${|\mathcal{P}| \choose 2}$} pairwise tests among the partitions in $\mathcal{P}$ satisfies the null hypotheses, the vertices corresponding to $\mathcal{P}$ would ideally form the largest clique in graph $G$.
However, this ideal situation is unlikely to happen.
Instead, due to the probabilistic nature of our 
theoretical guarantee in \eqref{eq:stepdown}, there are likely to be a few missing edges in $G$ among the 
vertices corresponding to $\mathcal{P}$.
Hence, a natural task is to instead find the largest \emph{quasi-clique}, a task that has been well-studied
by the computer science community (see \cite{tsourakakis2014novel} and its references within). 
We say a set of $k$ vertices form a $\gamma$-quasi-clique if there are 
at least $\gamma \cdot {k \choose 2}$ edges among these $k$ vertices for some $\gamma \in [0,1]$. 
The largest $\gamma$-quasi-clique is the largest vertex set that forms
a $\gamma$-quasi-clique . 
We justify the choice to search for this $\gamma$-quasi-clique
since, by our model \eqref{eq:hiermodel}, the prevalent covariance matrix
among the $r$ partitions is the desired covariance matrix $\bSigma$ we wish to estimate.
Here, $\gamma$ is an additional tuning parameter, but we set $\gamma =0.95$ by default throughout this
entire paper.

\begin{figure}[tb]
  \centering
  \includegraphics[width=400px]{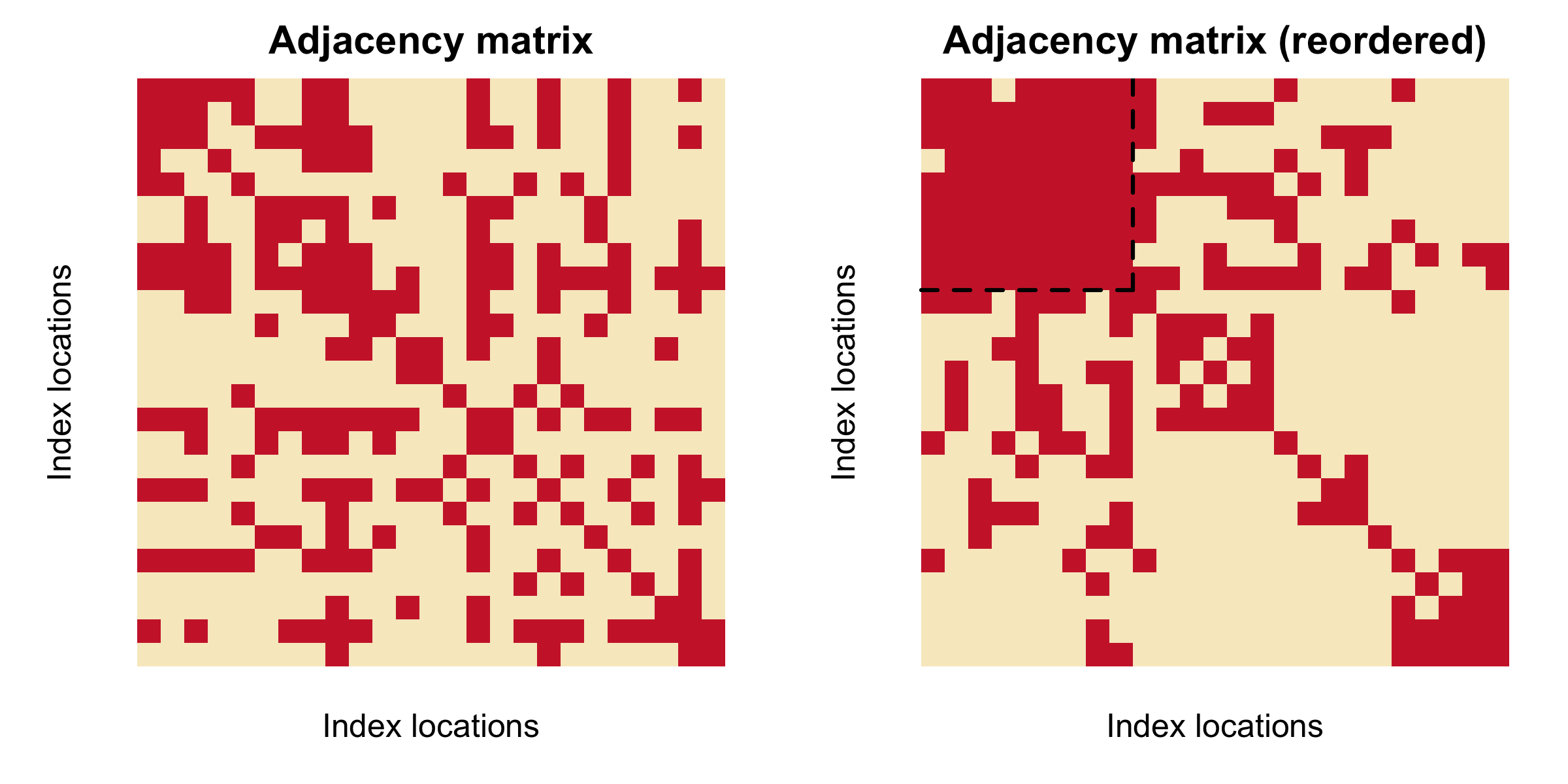}
    \vspace{-1em}
  \caption
   {\small (A) Visualization of an (example) adjacency matrix that can be formed using \eqref{eq:hypothesis_graph}, where the $i$th row from top and column from the left denotes the
   $i$th vertex. A red square in position $(i,j)$ denotes an edge between vertex $i$ and $j$, and a pale square
   denotes the lack of an edge.
   (B) Illustration of the desired goal. The rows and columns
   are reordered from Figure A, and the dotted box denotes the vertices that were found 
   to form a $\gamma$-quasi-clique.}
    \label{fig:goal}
\end{figure}

Unfortunately, many algorithms that could be used to find the largest $\gamma$-quasi-clique do not
satisfy a certain monotone property in practice, which hinders their usability.
Specifically, consider an algorithm $\mathcal{A}$ that takes in
a graph $G$ and outputs a vertex set, denoted by $\mathcal{A}(G)$, 
and for two graphs $G'$ and $G$, let $G' \subseteq G$ denote that $G'$ is a subgraph of $G$.
We say that algorithm $\mathcal{A}$ has the \emph{monotone property} if
\begin{equation} \label{eq:property}
G' \subseteq G \quad\Rightarrow\quad |\mathcal{A}(G')| \leq |\mathcal{A}(G)|, \quad \text{for any two graphs }G, G'.
\end{equation}
We are not aware of such a property being important in the quasi-clique literature, but it is a natural property to inherit from the multiple testing community.
That is, a multiple testing procedure has the monotone property if increasing the signal-to-noise ratio (i.e., decreasing the p-values) yields more rejections
(see \citep{hahn2018closure} and references within).
Similarly in the quasi-clique setting, 
it is natural to expect that increasing the signal-to-noise ratio (i.e., removing edges in $G$)
yields less partitions selected.
The monotone property is crucial in practice since 
it can be shown that the chosen FWER level $\alpha$ 
and the graph $G$ defined in \eqref{eq:hypothesis_graph}
have the following relation,
\[
\alpha \geq \alpha' \quad\Rightarrow\quad  G \subseteq G',
\]
where $G$ and $G'$ are the graphs formed by FWER level $\alpha$ and $\alpha'$ respectively.
Hence, 
an algorithm that does not exhibit the property in \eqref{eq:property}
will be fragile -- using a smaller $\alpha$ to accept more null hypotheses might 
counterintuitively result in less partitions being selected.
As we will demonstrate in \Fref{sec:simulation} through simulations, many existing algorithms to find the largest
quasi-clique do not satisfy the monotone property empirically. Therefore, we develop the following new algorithm
to remedy this. 

We describe the algorithm below. It starts by constructing a list containing 
all maximal cliques in the graph
based on  \eqref{eq:hypothesis_graph}. A maximal clique is a vertex set that forms a clique but
is not subset of a larger clique. The algorithm then proceeds by determining if the union of any
two vertex sets forms a $\gamma$-quasi-clique. If so, this union of vertices is added to the
list of vertex sets. The algorithm returns the largest vertex set in the list
when all pairs of vertex sets are tried and no new
$\gamma$-quasi-clique is found. 
We demonstrate in \Fref{sec:simulation} that this algorithm exhibits the monotone 
property \eqref{eq:property} empirically.

\vspace{1em}
\textbf{\labelText{Algorithm 4}{label:alg4}: Clique-based selection}
\vspace{-1em}
\begin{enumerate}
\item Form graph $G$ based on \eqref{eq:hypothesis_graph}.
\item Form $\mathcal{Q}$, the set of all vertex sets that form a
maximal clique in $G$. Each vertex set is initialized with a child set equal to itself.
\item While there are  vertex sets $A, B \in \mathcal{Q}$ the algorithm has not
tried yet:
\begin{enumerate}
\item Determine if $C = A \cup B$ forms a
$\gamma$-quasi-clique in $G$. If so, add $C$ as a new vertex set into $\mathcal{Q}$, with $A$ and $B$ 
as its two children sets.
\end{enumerate}
\item Return the largest vertex set in $\mathcal{Q}$.
\end{enumerate}

\vspace{1.5em}
\begin{figure}
  \centering
  \includegraphics[width=300px]{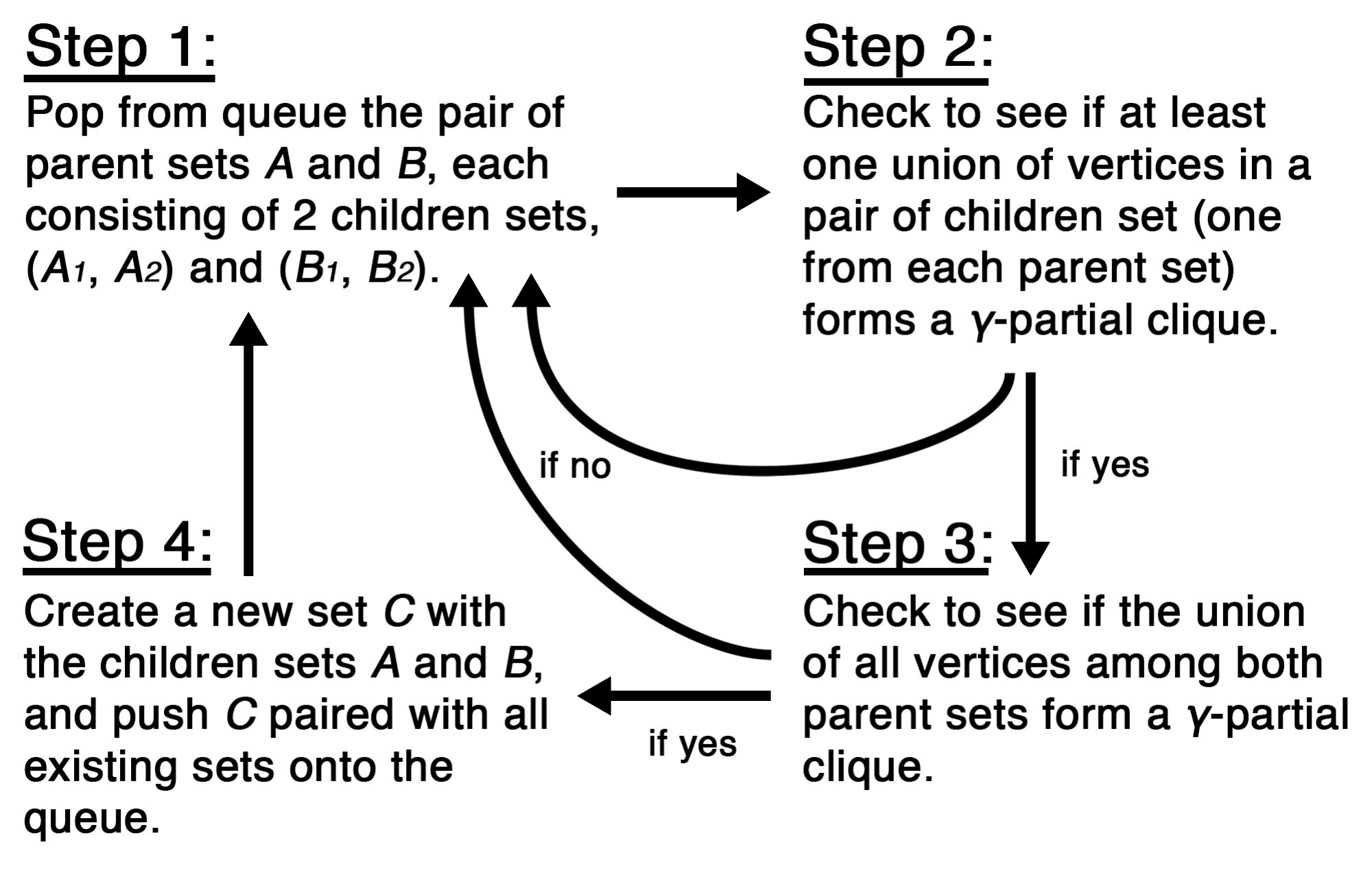}
  \caption
   {\small Schematic of \nameref{label:alg4}'s implementation. Step 2 is able to leverage hash tables which stores
   previous calculations to see if the union of vertices in a pair of children sets forms a $\gamma$-quasi-clique.
   This has a near-constant computational complexity. This can save tremendous computational time since
   Step 3, which checks if the union of vertices in both parent sets form a $\gamma$-quasi-clique, has a computational
   complexity of $O(r^2)$.}
    \label{fig:queue}
\end{figure}

A naive implementation of the above algorithm would require
checking if an exponential number of vertex set unions $C = A \cup B$ forms
a $\gamma$-quasi-clique, and each check requires $O(r^2)$ operations.
However, we are able to dramatically reduce the number of checks required by using the following heuristic:
we only check whether the union of $A$ and $B$ forms a $\gamma$-quasi-clique
if the union of two children sets, one from each $A$ and $B$, forms a $\gamma$-quasi-clique.
This heuristic allows us to exploit previous calculations and reduce computational costs.
We implement this idea by using
 one hash table to record which vertex sets are children of other vertex sets, and another
hash table table to record if the union of two vertex sets forms a $\gamma$-quasi-clique.
This idea is illustrated in \Fref{fig:queue}. 
Additional details on how to initialize and optionally post-process \nameref{label:alg4} are given in Appendix \ref{sec:quasiclique}.

\section{Simulation study} \label{sec:simulation}

We perform empirical studies to show that COBS
has more power and yields a better estimation of the desired covariance matrix $\bSigma$
over conventional methods as
the samples among different partitions are drawn from increasingly dissimilar distributions. 

\textbf{Setup}:
We generate synthetic data in $r=25$ partitions, where the data in each partition has 
$n=15$ samples and $d=1000$ dimensions drawn from a non-Gaussian distribution.
Among these $r$ partitions,
the first group of $r_1=15$ partitions, second group of $r_2 = 5$ partitions and third group of $r_3=5$ partitions
are drawn from three different nonparanormal distributions respectively \citep{liu2009nonparanormal}. 
The goal in this simulation suite is to detect these $r_1$ partitions with the same covariance structure.
The nonparanormal distribution is
a natural candidate to 
model genomic data with heavier tails and multiple modes 
 (\cite{liu2012high} and \cite{xue2012regularized}), and serves to demonstrate that our methods in 
  \Fref{sec:method} does not rely on the Gaussian assumption.
Formally, a random vector $\bX = (X_1, \ldots, X_d) \in \mathbb{R}^d$ is drawn from a nonparanormal distribution if there exists $d$
monotonic and differentiable functions $f_1,\ldots,f_d$ such that when applied marginally, $\bZ = (f_1(X_1), \ldots, f_d(X_d)) \sim N(\bmu, \bSigma)$, a Gaussian distribution with proxy mean vector $\bmu$ and proxy covariance matrix\footnote{We emphasize ``proxy'' covariance matrix, for example, since the covariance of $\bX$, the random variable we sample, is not $\bSigma$.} $\bSigma$.
We provide the details of how we generate the three nonparanormal distributions in Appendix \ref{sec:simulation_app},
but we highlight the key features regarding $\bSigma$ below.

\begin{figure}[tb]
  \centering
  \includegraphics[width=400px]{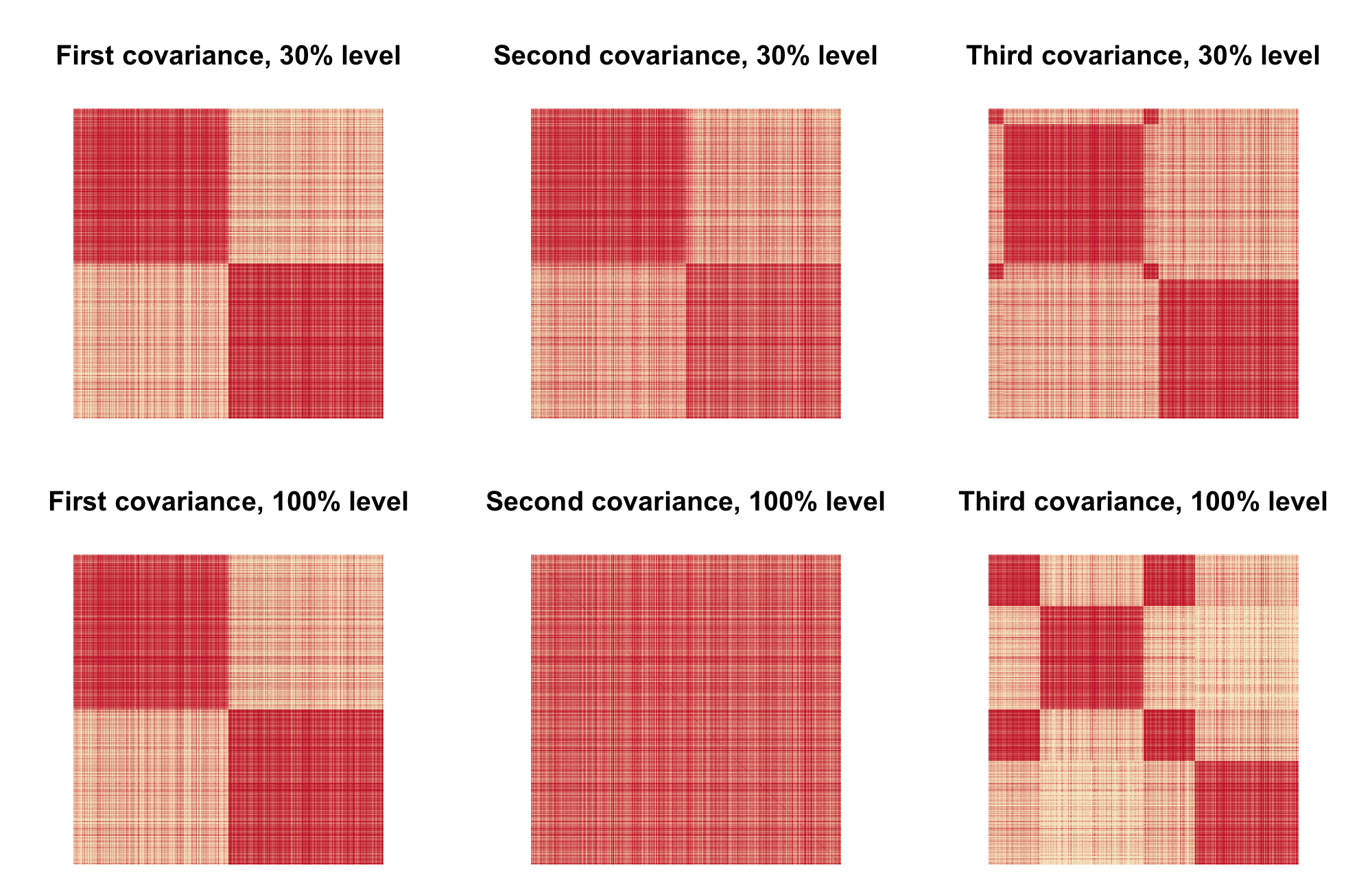}
  \caption
   {\small (Top row) Heatmap visualizations of the empirical covariance matrix of the three partitions, each drawn from
   a different nonparanormal distribution when $\beta = 0.3$.
   The distribution using $\bSigma^{(1)}$, $\bSigma^{(2)}$ and $\bSigma^{(3)}$ are 
   shown as the left, middle and right plots respectively.
   The darker shades of red denote a higher covariance.
   (Bottom row) Visualizations similar to the top row except $\beta =1$, so the dissimilarity comparing $\bSigma^{(2)}$ or $\bSigma^{(3)}$ to $\bSigma^{(1)}$ is increased.}
    \label{fig:covariance}
\end{figure}

We construct three different proxy covariance matrices $\bSigma^{(1)}, \bSigma^{(2)},$ and $\bSigma^{(3)}$ in such
a way that for a given parameter $\beta \in [0,1]$, we construct $\bSigma^{(2)}$ and $\bSigma^{(3)}$ to be more
dissimilar from $\bSigma^{(1)}$ as $\beta$ increases.
 We highlight the key features of our constructed proxy covariance matrices here.
All three proxy covariance matrices are all based on a stochastic block model (SBM),
a common model used to model gene networks \citep{liu2018global,funke2019stochastic}.
The first $r_1$ partitions are generated using proxy covariance matrix $\bSigma^{(1)}$,
which is an SBM with two equally-sized clusters where the within-cluster covariance is $a = 0.9$ and the between-cluster covariance is $b=0.1$. 
The second $r_2$ partitions  are generated using proxy covariance matrix $\bSigma^{(2)}$, which is
similar to $\bSigma^{(1)}$ except $a$ and $b$ are shrunk towards $0.5$ depending on the magnitude of $\beta$. 
The last $r_2$ partitions  are generated using proxy covariance matrix $\bSigma^{(3)}$, which is
similar to $\bSigma^{(1)}$ except an equal fraction of variables from both clusters break off to form a third cluster, depending on the magnitude of $\beta$.
 By generating $\bSigma^{(1)}, \bSigma^{(2)}$, and $\bSigma^{(3)}$ in this fashion, the parameter $\beta$ can control the difficulty of the simulation setting -- 
 a larger $\beta$ means COBS would ideally have more power in distinguishing 
among the first $r_1$ partitions from the other partitions. 
\Fref{fig:covariance} visualizes the resulting covariance matrices for the three nonparanormal distribution we generate
in this fashion for $\beta=0.3$ and $\beta=1$.

\textbf{Multiple testing}:
We use the Stepdown method described in \Fref{ss:test} on our simulated data
where $\beta = \{0, 0.3, 0.6, 1\}$ to see how the true positive rates and false positive rates vary
with $\beta$.
Let $\mathcal{L} = \{(i_1, j_1), (i_2, j_2), \ldots\}$ denote the returned set of partition pairs that correspond to the accepted
null hypothesis.  
Since our goal is to find the first $r_1$ partitions, we define the true positive rate and false positive rate
for individual hypotheses to be
\begin{align*}
\text{True positive rate (TPR) for hypotheses} = \frac{\big|\big\{(i,j) \in \mathcal{L} : i \leq r_1 \text{ and } j \leq r_1\big\}\big|}{{r_1 \choose 2}},\\
\text{False positive rate (FDR) for hypotheses} = \frac{\big|\big\{(i,j) \in \mathcal{L} : i  > r_1 \text{ or } j > r_1\big\}\big|}{{r \choose 2} - {r_1 \choose 2}}.
\end{align*}

\begin{figure}[tb]
  \centering
  \includegraphics[width=400px]{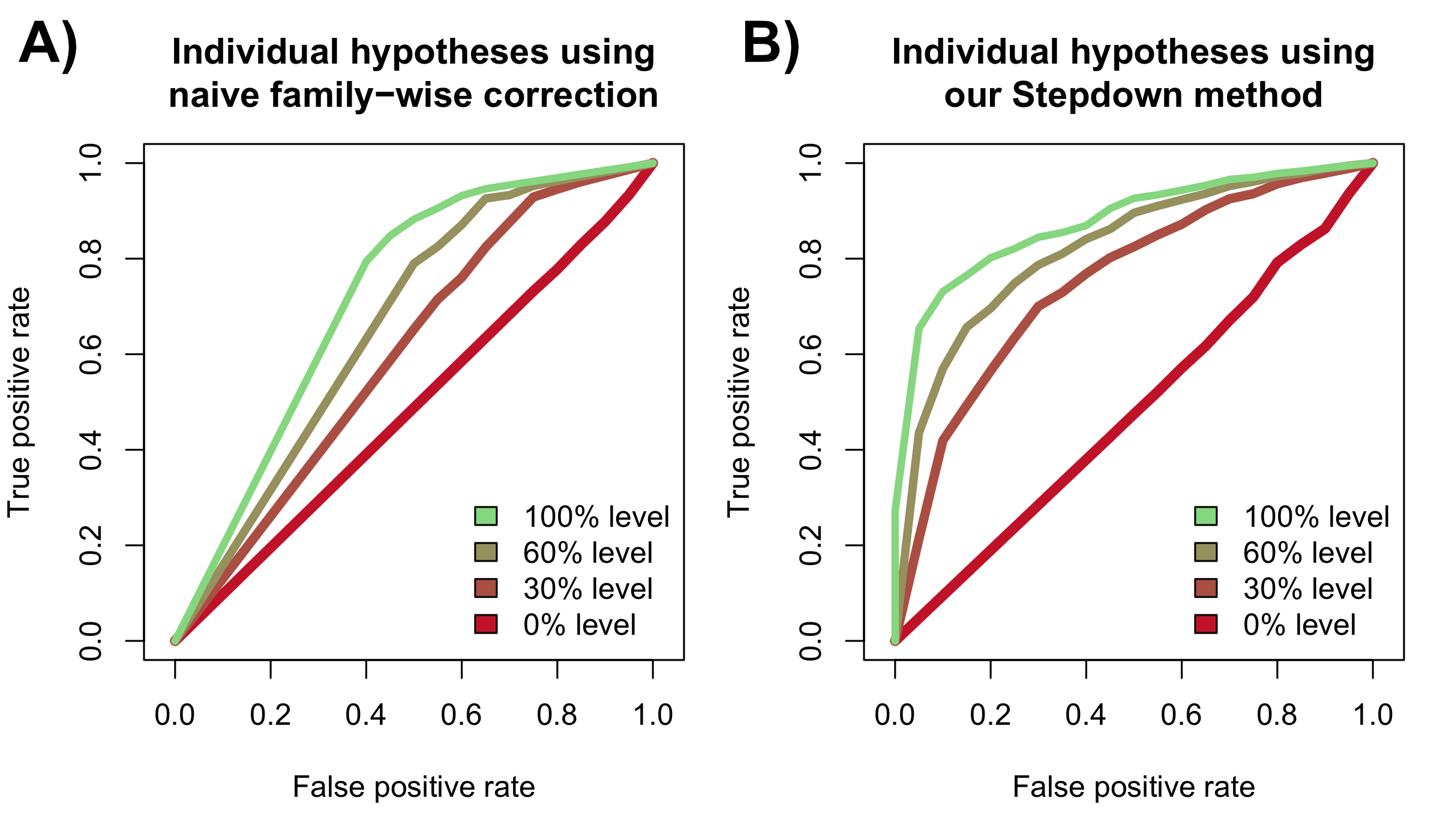}
    \vspace{-1em}
  \caption
   {\small RoC curves for the accepted null hypotheses, for settings where $\beta = (0, 0.3, 0.6, 1)$,
   where each curve traces out the results as $\alpha$ varies from 0 to 1. (A) The curves resulting from using a Bonferroni correction to the ${r \choose 2}$ individual hypothesis tests. 
   (B) The curves resulting from using our Stepdown method. 
   }
    \label{fig:roc_hypothesis}
\end{figure}

We plot the RoC curves visualizing the TPR and FPR in \Fref{fig:roc_hypothesis}. Each curve traces out the mean true and false positive rate over 25 simulations as $\alpha$ ranges from $0$ (top-right of each plot) to $1$ (bottom-left of each plot), where we use 200 bootstrap trials per simulation. 
\Fref{fig:roc_hypothesis}A shows the naive analysis where
we compute all ${r \choose 2}$ p-values, one for each hypothesis test comparing two partitions, and accept hypotheses for varying levels of $\alpha$ after using a Bonferroni correction. 
\Fref{fig:roc_hypothesis}B shows the Stepdown method.
In both plots, we see that as $\beta$ increases, each method 
has more power.
However, as we mentioned in \Fref{ss:test}, there is a considerable loss of power when comparing
the Bonferroni correction to the Stepdown method. This is because the Bonferroni correction is too conservative when accounting for dependencies.

\textbf{Partition selection}:
After using Stepdown,
we proceed to select
the partitions as in \Fref{ss:selection} to understand the
monotone property and see how the true and false positive rates for partitions vary with $\beta$.

\begin{figure}[tb]
  \centering
  \includegraphics[width=400px]{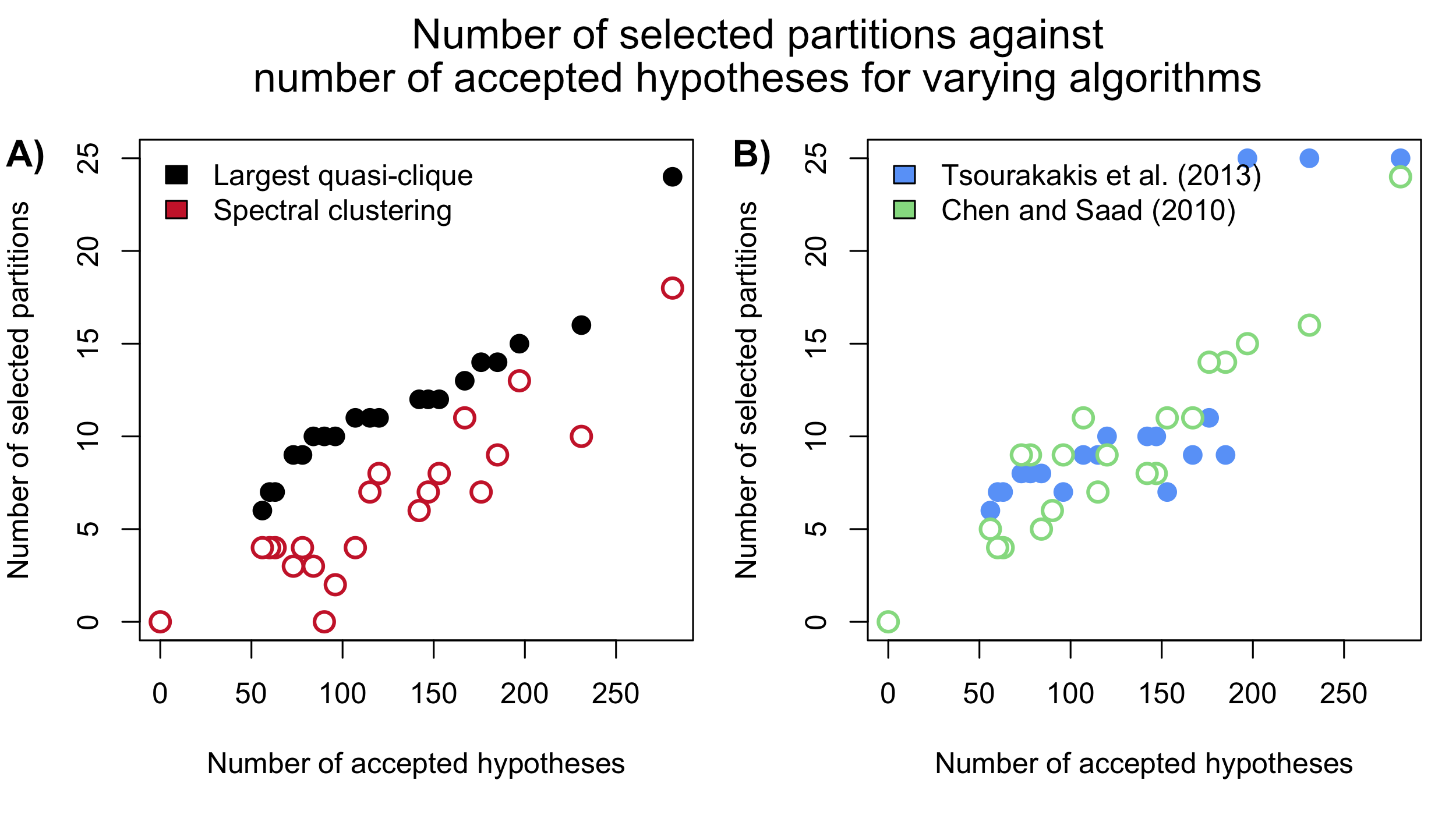}
    \vspace{-0.5em}
  \caption
   {\small Number of selected partitions for a particular simulated dataset as the number of
   accepted null hypotheses varies with the  FWER level $\alpha$. 
   (A) Results using our clique-based selection method developed in \Fref{ss:selection} and spectral clustering.
   (B) Results using the methods developed in \cite{tsourakakis2013denser} and  \cite{chen2010dense}. See
    Appendix \ref{sec:quasiclique} for more details of these methods.
   }
    \label{fig:num_partitions}
\end{figure}

\Fref{fig:num_partitions} shows that three 
methods currently in the literature that can be used to find the largest quasi-clique in
 \eqref{eq:hypothesis_graph} fail the monotone property \eqref{eq:property}, whereas COBS succeeds.
In  \Fref{fig:num_partitions}A, we compare our clique-based selection method, described in \Fref{ss:selection}, 
against spectral clustering, a method used in network analyses designed to find highly connected vertices
\citep{lei2015consistency}, whereas in \Fref{fig:num_partitions}B, two methods recently developed in the 
computer science community are compared (\cite{chen2010dense} and \cite{tsourakakis2013denser}). These three methods are detailed in
Appendix \ref{sec:quasiclique}, and all the methods 
receive the same set of accepted null hypotheses as the FWER level $\alpha$ varies.
Recall that since the Stepdown method accepts more hypotheses as $\alpha$ decreases,
the graph formed by \eqref{eq:hypothesis_graph} becomes denser as $\alpha$ increases. 
However, as we see in \Fref{fig:num_partitions}, the number of partitions selected by all but our
method sometimes decreases as number of accepted null hypotheses increases, hence violating the desired
monotone property.

\begin{figure}[tb]
  \centering
  \includegraphics[width=425px]{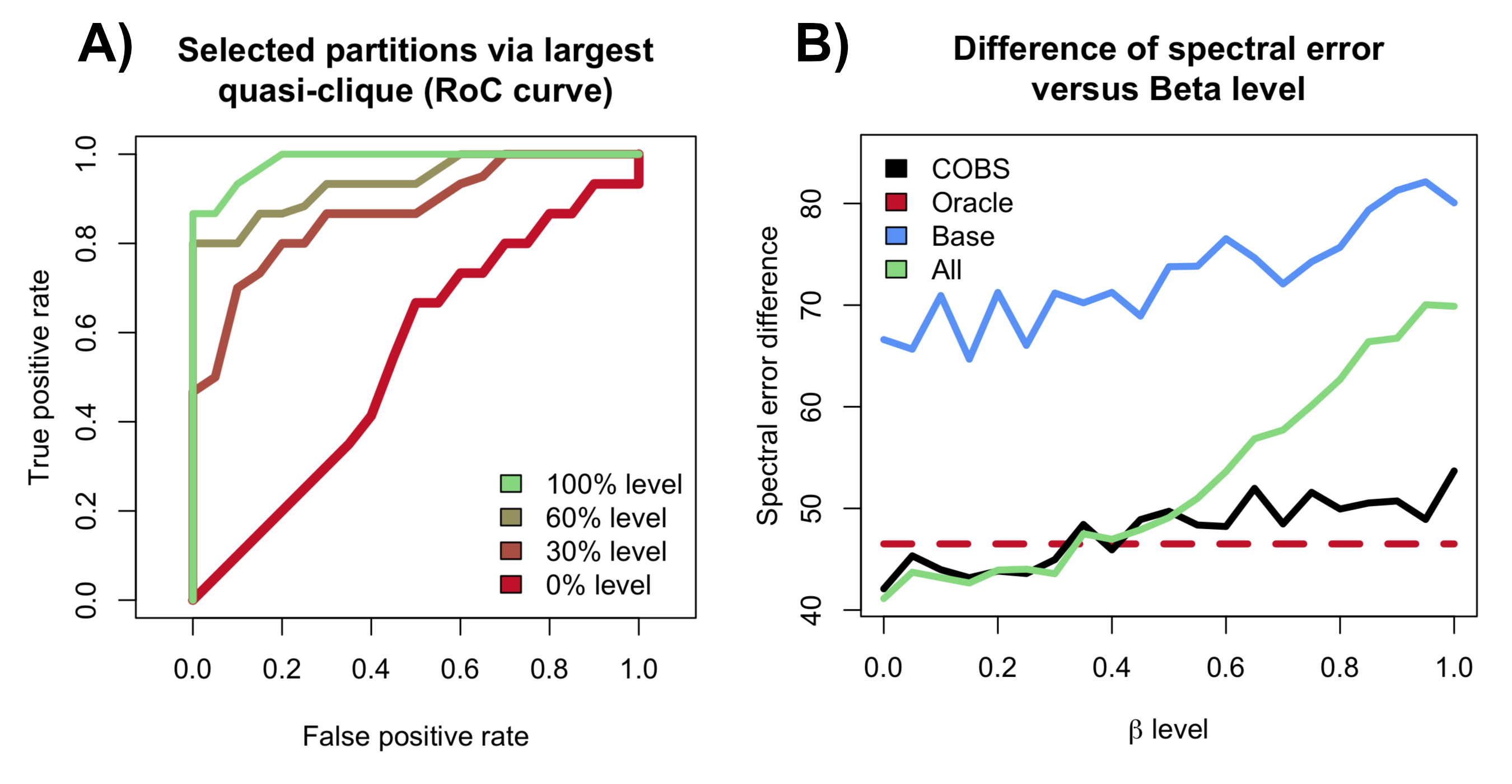}
    \vspace{-1em}
  \caption
   {\small A) Similar RoC curves to \Fref{fig:roc_hypothesis}, but for selected partitions selected by COBS.
   B) The mean spectral error of each method's downstream estimated covariance matrix
   for varying $\beta$ over 25 trials. The four methods to select partitions shown are COBS for $\alpha = 0.1$ (black), the method that selects all partitions (green), the method that selects a fixed
   set of 5 partitions (blue), and the method that selects exactly the partitions that contain samples
   drawn from a nonparanormal distribution with proxy covariance $\bSigma^{(1)}$ (red). }
    \label{fig:snr}
\end{figure}

\Fref{fig:snr}A shows the RoC curves for varying $\beta$ as the FWER level $\alpha$ varies. This figure is closely related to \Fref{fig:roc_hypothesis}B. 
We use our clique-based selection method to find the largest $\gamma$-quasi-clique for $\gamma=0.95$. Let $\hat{\mathcal{P}}$ denote the selected set of partitions. Similar to before, we define the TPR and FPR in this setting as
\begin{align*}
\text{TPR for partitions} &= \frac{\big|\big\{p \in \hat{\mathcal{P}}: p \leq r_1 \big\}\big|}{r_1},\\
\text{FDR for partitions} &= \frac{\big|\big\{p \in \hat{\mathcal{P}}: p > r_1 \big\}\big|}{r_2+r_3}.
\end{align*}
\noindent We see that the power of the COBS increases as $\beta$ increases, as expected.

\textbf{Covariance estimation}: Finally, we show that COBS is able to improve the downstream covariance estimation compared to other approaches. 
To do this, we use four different methods to select partitions and 
compute the empirical covariance matrix among the samples in those partitions.
The first three methods resemble analyses that could be performed on the BrainSpan dataset in practice.
The first method uses the COBS.
The second method always selects all the partitions, which resembles
using all the partitions in the BrainSpan dataset. 
The third method always selects the same 5 partitions -- 3 partitions contain samples drawn from 
the nonparanormal distribution with proxy covariance $\bSigma^{(1)}$, 
while the other 2 partitions contain samples from each of the remaining two distributions. 
This resembles previous work \citep{liu2014network} that consider only partitions in Window 1B.
For comparison, the last method resembles an oracle that
selects exactly the $r_1$ partitions containing samples drawn the nonparanormal distribution
 with proxy covariance $\bSigma^{(1)}$.

\Fref{fig:snr}B shows that our partition selection procedure performs
almost as well as the oracle method over varying $\beta$ level.
Notice that for low $\beta$, COBS and the method using all partitions yield a smaller spectral error than the oracle
method.  
This is because for low $\beta$, the covariance matrices $\bSigma^{(1)}$, $\bSigma^{(2)}$, and $\bSigma^{(3)}$ are
almost indistinguishable. However, as $\beta$ increases, the dissimilarities among $\bSigma^{(1)}$, $\bSigma^{(2)}$, and $\bSigma^{(3)}$ grow. This means methods that do not adaptively choose which partitions to select become increasingly worse. However, our procedure remains competitive,
performing almost as if it knew which partitions contain samples drawn the nonparanormal distribution
 with proxy covariance $\bSigma^{(1)}$.  
Additional simulations that go beyond the results in this section 
are deferred to Appendix \ref{sec:additional_sim}.

\vspace{-1.5em}
\section{Application on BrainSpan study} \label{sec:brainspan}

We demonstrate the utility of COBS by applying it within the DAWN framework established in
 \cite{liu2014network}. Specifically, in this section, we ask two questions. First, does COBS select
reasonable partitions within the BrainSpan data, given our current scientific understanding outlined in \Fref{sec:model}? 
Second,
does using COBS within the DAWN framework lead to a more meaningful gene co-expression
network that can implicate genes using a ``guilt-by-association'' strategy?

Here, we discuss the different datasets relevant to the analysis in this section. 
DAWN relies on
two types of data to identify risk genes: gene expression data to estimate a gene co-expression
network and genetic risk scores
to implicate genes associated with ASD. 
For the former, we use the
BrainSpan microarray dataset \citep{kang2011spatio}, which has been the primary focus of this article so far. 
For the latter, we use the TADA scores
published in \cite{de2014synaptic}
which are p-values, one for each gene, resulting from a test
for marginal associations with ASD based on rare genetic variations and mutations.\footnote{TADA stands for Transmission and De novo association \citep{he2013integrated}.}
For enrichment analysis, we use a
third dataset consisting of TADA scores from \cite{satterstrom2019large}.
We use this third dataset only to assess the quality of our findings, and
these TADA scores are derived as in \cite{de2014synaptic}, but include additional
data assimilated since 2014. 
Relying on a later “data freeze,'' this 2019 study has greater power
to detect risk genes compared to the 2014 study: the two studies report 102 and 33 risk genes, respectively, with FDR 
cutoff of $10\%$. 
Additional details of our analysis in this section can be found in Appendix \ref{sec:additional_analysis}.

\vspace{-1.5em}
\subsection{Gene screening} \label{ss:screening}
We first preprocess the BrainSpan data by determining which genes to
include in our analysis. This is necessary since there are over 13,939 genes in the BrainSpan dataset,
 most of
which are probably not correlated with any likely risk genes. Including such genes increases
the computationally cost and is not informative for our purposes.
Hence, we adopt a similar screening procedure as in \cite{liu2014network},
which involves first selecting genes with high TADA scores based on \cite{de2014synaptic}, 
and then selecting all genes with
a high Pearson correlation in magnitude 
with any of the aforementioned genes within the BrainSpan dataset.
We select a total of 3,500 genes to be used throughout the remainder of this analysis.

\subsection{Partition selection} \label{ss:partition}
Motivated by the findings in \cite{willsey2013coexpression},
we analyze the BrainSpan dataset using COBS to find many partitions
that are homogeneous with most partitions in Window 1B (\Fref{fig:partpreselect}).
We use the Stepdown method 
with 200 bootstrap trials and FWER level $\alpha = 0.1$.
This simultaneously finds which null hypotheses are accepted
among the ${125 \choose 2}$ hypotheses tested. 
Based on these results,
we select the partitions that
form the largest $\gamma$-quasi-clique for $\gamma = 0.95$.

\begin{figure}
  \centering
  \includegraphics[width=375px]{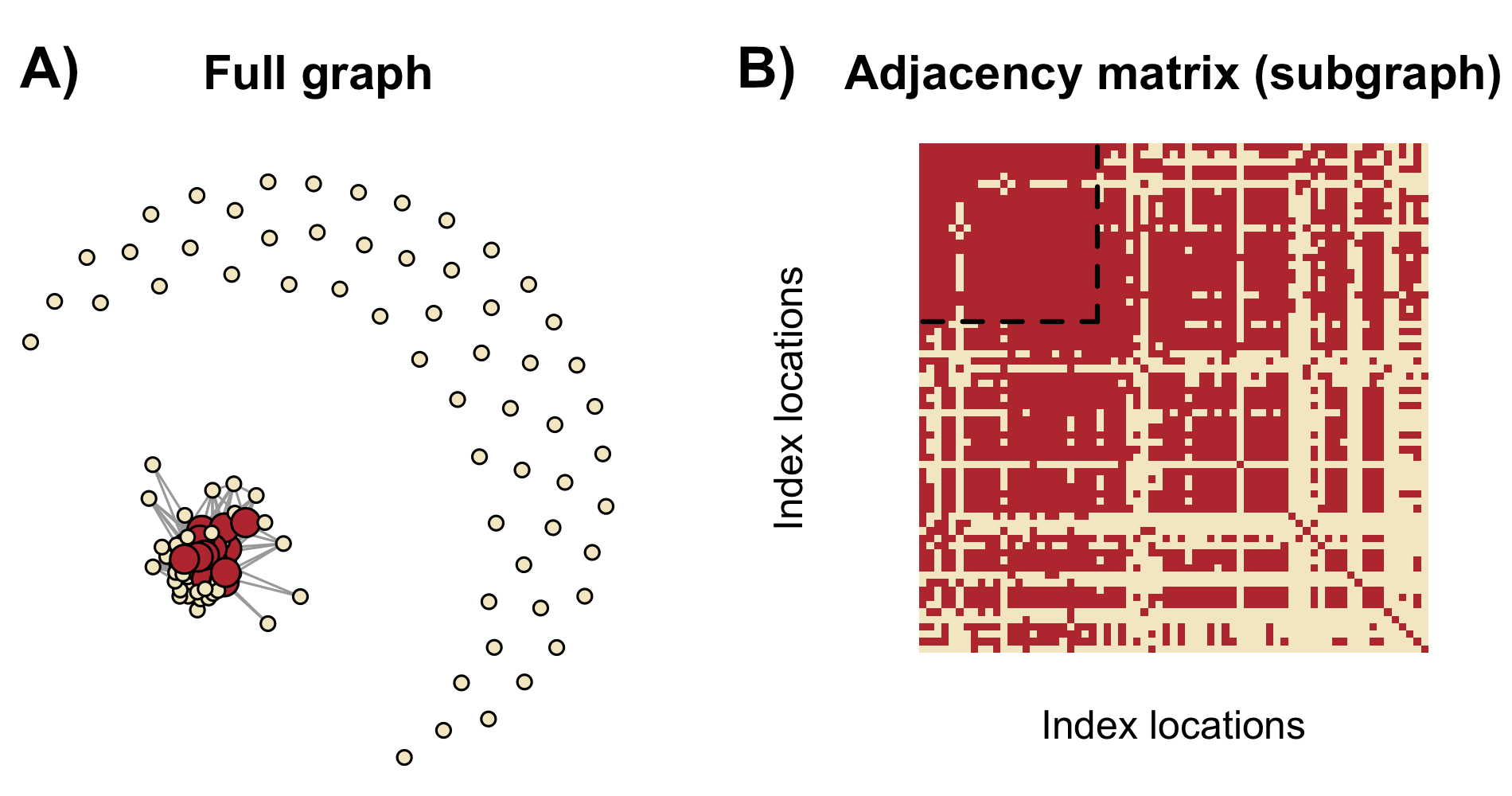}
    \vspace{-1em}
  \caption
   {\small (A) The graph $G$ containing all 125 nodes.
   Red nodes correspond to the 24 selected partitions, while pale nodes correspond to partitions not
   selected.  (B) The adjacency matrix of a connect component of $G$, where each row and corresponding 
   column represents a different node, similar to \Fref{fig:goal}. A red pixel corresponds to an edge between
   two nodes, while a pale pixel represents no edge.}
    \label{fig:part}
\end{figure}

We visualize the results of the Stepdown method in \Fref{fig:part}, illustrating that COBS finds 24 partitions which 
have statistically indistinguishable covariance matrices, 7 of which are in Window 1B. 
We form the graph $G$ based on the accepted null hypotheses, as described in \eqref{eq:hypothesis_graph}.
\Fref{fig:part}A shows the full graph with all 125 nodes,
while \Fref{fig:part}B shows the connected component of $G$ as an adjacency matrix.
We can
see that the 24 partitions we select, which contain 272 microarray samples, correspond to 24 nodes in $G$ that form a dense quasi-clique.

\begin{figure}
  \centering
  \includegraphics[width=450px]{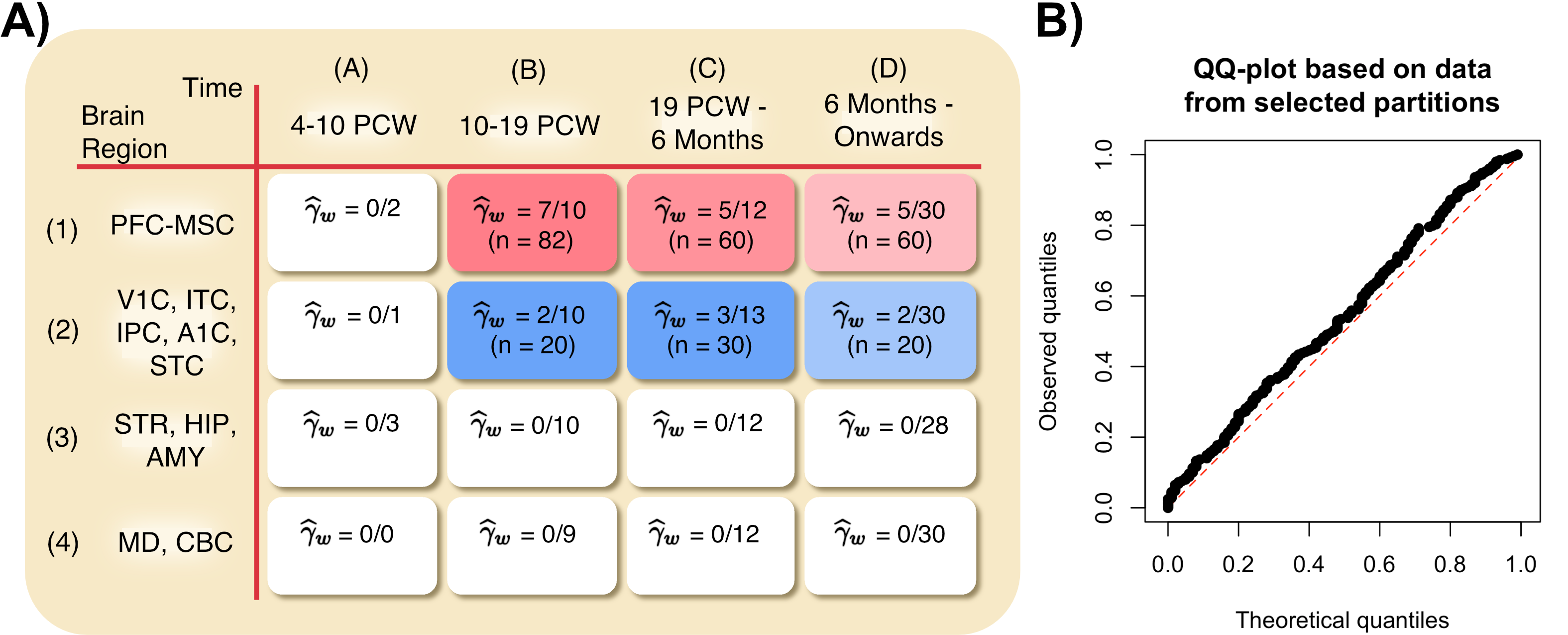}
    \vspace{-0.5em}
  \caption
   {\small (A) The number of partitions and samples ($n$) selected within each
   window. Partitions from 6 different windows are chosen,
   and the estimated $\gamma_w$ is empirical fraction of selected
   partitions within each window. The more vibrant colors
display a higher value of $\hat{\gamma}_w$. 
 (B) A QQ-plot of the 250 p-values generated when applying our diagnostic to the 24
   selected partitions, similar to \Fref{fig:initial_qq}. While these p-values are slightly
   left-skewed, they suggest that the selected partitions
   are more homogeneous when compared to their counterparts shown in \Fref{fig:initial_qq}.}
    \label{fig:selected_partitions}
\end{figure}

We visualize the proportion of selected partitions per window in the BrainSpan
dataset in \Fref{fig:selected_partitions}A
to demonstrate that our findings are consistent with the findings in
\cite{willsey2013coexpression}. As mentioned in \Fref{sec:model}, 
\cite{willsey2013coexpression} find that partitions in Window 1B are mostly
homogeneous and are enriched for tightly clustered risk genes.
The authors also found that, on average, gene expression varies
smoothly across developmental periods, meaning there is greater
correlation between the gene expressions belonging to adjacent developmental windows.
The authors also estimate a hierarchical clustering among the four brain regions.
Indeed, our results match these finding. We select a large proportion of partitions in Window 1B,
and the proportion of selected partitions smoothly decreases as the window representing older developmental
periods as well as brain regions become more dissimilar to Window 1B. 

Lastly, we apply the same diagnostic as in \Fref{sec:elementary} to show in \Fref{fig:selected_partitions}B that the 272
samples within our 24 selected
partitions are much more homogeneous than the 107 samples among the 10 partitions in Window 1B.
The p-values we obtain after 250 divisions are much closer to uniform that those shown in \Fref{fig:initial_qq}.

\subsection{Overview of DAWN framework} \label{ss:gene_network}

\begin{figure}
  \centering
  \includegraphics[width=425px]{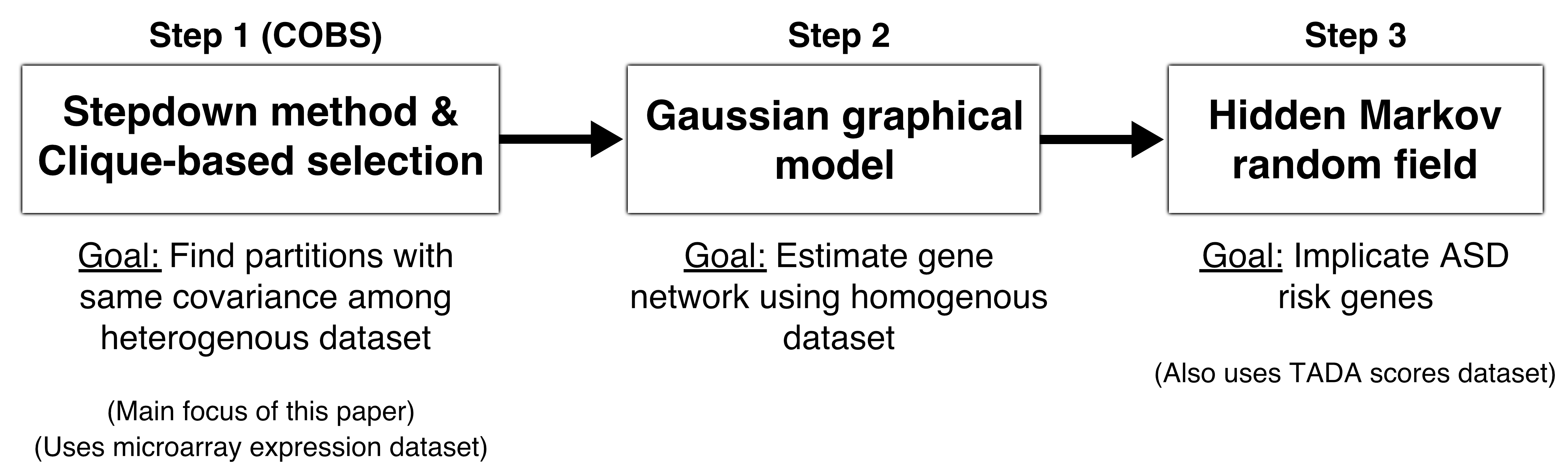}
  \caption
   {\small Flowchart of how COBS (Stepdown method and clique-based selection method)
   is used downstream to find risk genes within the DAWN framework. Step 2 and 3 are taken directly from \cite{liu2014network}.}
    \label{fig:dawn_flowchart}
\end{figure}

As
alluded to in \Fref{sec:intro}, DAWN estimates a gene co-expression network using the microarray
partitions to boost the power of the TADA scores using a “guilt-by-association” strategy. 
\Fref{fig:dawn_flowchart} illustrates this procedure as a flowchart.
The first step uses COBS
to select 24 partitions from the BrainSpan dataset, as stated in the previous subsection.
In the second step,
DAWN estimates a Gaussian graphical model
via  neighborhood selection \citep{meinshausen2006high}  from the 272 samples in these partitions
to represent the gene co-expression network. 
In the third step, DAWN implicates risk genes via
a Hidden Markov random field (HMRF) model by combining the Gaussian graphical model with the
TADA scores.
The details are in \cite{liu2014network}, but in short, this procedure
assumes a mixture model of the TADA scores between risk genes and non-risk genes,
and the probability that a gene is a risk gene depends on the graph structure.  
An EM algorithm is used to estimate the parameters of this HMRF model, after which
a Bayesian FDR procedure \citep{muller2006fdr} is used on the estimated posterior probabilities of being a risk gene
to output the final set of estimated risk genes.
The methodology in the second and third step are the same as those in \cite{liu2014network}, as we wish 
to compare only different ways to perform the first step.

\subsection{Investigation on gene network and risk genes} \label{ss:val}

In this subsection, we compare the 
DAWN analysis using the 24 partitions selected by COBS (i.e., the ``COBS analysis'')
to using the 10 partitions in Window 1B
originally used in \cite{liu2014network} (i.e., the ``Window 1B analysis'') to show
how COBS improves the estimated gene network.

\paragraph{Closeness of genes within co-expression network.}
We demonstrate that the 102 genes detected by the 
newer  TADA scores \citep{satterstrom2019large} are roughly 10\%-30\% 
closer to the 33 genes detected
by the older TADA scores \citep{de2014synaptic} 
in the gene network estimated in the COBS analysis
than in the Window 1B analysis. 
This suggests that the COBS analysis estimates a more useful gene network,
because when future TADA scores are published after \cite{satterstrom2019large}, the next wave of
detected risk genes are more likely to also be closer to the original risk genes detected in \cite{de2014synaptic}.
We defer the details to Appendix \ref{sec:additional_analysis}, but highlight the procedure to derive
this result here.
Effectively comparing the distances between genes in a network is a difficult problem since the 
estimated gene networks in the COBS and Window 1B analyses have different number of edges.
In addition, current research has suggested that
natural candidates such as the shortest-path distance or the commute distance do not accurately
capture the graph topology (\cite{alamgir2012shortest} and \cite{von2014hitting}). 
Hence, we use two different distance metrics to measure the closeness of two sets of genes
that potentially overcome this problem. The first is using the path distance via the minimum spanning tree,
and
the second is using the Euclidean distance via the graph root embedding \citep{lei2018network}. 
Using either of these metrics lead to the same conclusion.

\paragraph{Enrichment analysis.} 
We demonstrate that COBS improves DAWN's ability to predict risk genes
based on the newer TADA scores \citep{satterstrom2019large} when utilizing the older TADA scores \citep{de2014synaptic} as input.
Specifically, the COBS analysis and the Window 1B analysis implicate 209 and 249 risk genes 
respectively an FDR cutoff of 10\%, respectively. 
The risk genes implicated in the COBS analysis have a better enrichment for the 102 genes detected using
the newer TADA scores \citep{satterstrom2019large}: 18.8\% (COBS analysis) versus 14.6\% (Window 1B analysis).
We note that genes implicated by DAWN but not by the TADA scores are not considered false positives.
In fact, \cite{he2013integrated} suggests that there are upwards of 500 to 1000 genes that increase risk for ASD. 
Hence, we are unlikely to detect all of the true risk genes based on tests that rely on rare genetic variation alone.

\paragraph{Robustness to $\gamma$.} 
We additionally verify the robustness of the above enrichment results to the parameter $\gamma$.
Recall that $\gamma$ controls the density of the 
edges in the quasi-clique, as introduced in \Fref{ss:selection}, and we typically set $\gamma =0.95$ by default.
When we re-run the entire analysis with different values of $\gamma$ varying from 0.85 to 0.97 at intervals of 0.01, we 
obtain 13 different sets of estimated risk genes.
We stop at $\gamma = 0.97$ since larger values result in no partitions selected outside of Window 1B.
When we intersect all 13 sets of risk genes together, we find that 144 risk genes are implicated regardless
of the value of $\gamma$, of which 22.9\% are in the list of 102 risk genes found using only the newer TADA scores \citep{satterstrom2019large}.
This is a promising result, as it demonstrates that the implicated risk genes 
in the COBS analysis are more enriched than those in the Window 1B analysis for a wide range of $\gamma$.

\section{Conclusion and discussions} \label{sec:conclusion}

In this article, we develop COBS to select many partitions with statistically indistinguishable 
covariance matrices in order to better estimate graphical models
for ASD risk gene detection. Our procedure first applies a Stepdown method to simultaneously 
test all ${r \choose 2}$ hypotheses, each testing whether or not a pair of partitions
share the same population covariance matrix. The Stepdown method is critical since it
can account for the dependencies among all ${r \choose 2}$ hypotheses via bootstrapping the
joint null distribution. Then, our procedure uses a clique-based selection method to 
select the partitions based on the accepted null hypotheses. The novelty in this latter method 
is its ability to preserve monotonicity, a property stating that less partitions should be selected as
the number of accepted null hypotheses is smaller. 
We demonstrate empirically that the COBS achieves this property while common methods
such as spectral clustering do not.
When we apply COBS to the BrainSpan dataset, we find
scientifically meaningful partitions based on the results in \cite{willsey2013coexpression}.
We also find that COBS aids in clustering the risk genes detected
in \cite{satterstrom2019large} closer to the risk genes detected in \citep{de2014synaptic}
within the estimated gene co-expression network
and in getting a better enrichment in implicated risk genes via the DAWN analysis.

The theoretical role of the FWER level $\alpha$ is not well understood mathematically.
Specifically, 
while \eqref{eq:stepdown} provides a theoretical guarantee about the set of null hypothesis
accepted, we would like to prove a theoretical guarantee about the set of selected
partitions $\hat{\mathcal{P}}$. 
Towards this end, we suspect that with
some modification to COBS, closed testing offers
a promising theoretical framework (see \cite{dobriban2018flexible} and references within).
This will be investigated in future work.

COBS is applied directly to help implicate risk genes for ASD, but this line of work
has broader implications in genetics.
Due to the improvement of high throughput technologies, it has become increasingly accessible
to gather large amounts of gene expression data. This includes both microarray and RNA sequencing data.
However, as we have seen in this article, gene expression patterns can vary wildly
among different tissues.
Hence, it is challenging to select samples that are relevant for specific scientific tasks.
Beyond analyzing brain tissues, \cite{greene2015understanding} develop procedures
to select relevant samples amongst a corpus of microarray expression data
to estimate gene co-expression networks for different tissue types.
While \cite{greene2015understanding} does not motivate their method from a statistical model, 
our work provides a possible statistical direction for this research field to move towards.

\vspace{1em}
\noindent \textbf{Acknowledgments:} We thank Bernie Devlin and Lambertus Klei for the
insightful discussions about our analysis and results. We thank Li Liu and Ercument
Cicek
for providing the code used in \cite{liu2014network} to build off of. We also thank the anonymous
reviewers for helpful suggestions on how to restructure the simulations and analyses.

\bibliographystyle{apalike}
\bibliography{bib}

\newpage
\appendix

 \begin{center}
    {\LARGE\bf Appendix for ``Covariance-based sample selection for heterogeneous data: 
Applications to gene expression and autism risk gene detection''}
\end{center}

\section{Code and dataset} \label{sec:code}
\if0\blind{
The \texttt{R} code for replicating all analyses and figures in this article are hosted on GitHub 
in the repository \url{https://github.com/linnylin92/covarianceSelection}. 
The three major datasets used in this article are also included in the repository.
The first dataset is the BrainSpan microarray samples collected by \citep{kang2011spatio}. 
While the original dataset 
is publicly available on GEO (\url{https://www.ncbi.nlm.nih.gov/geo/query/acc.cgi?acc=GSE25219}),
we provide a locally preprocessed dataset, which was created to be amendable for our analysis in \texttt{R}.
The second dataset is the older TADA scores \citep{de2014synaptic}.
The third dataset is the list of 102 risk genes detected using the newer TADA scores \citep{satterstrom2019large}.
 }\fi
 
 \if1\blind{
 The details for the code for the method, simulation, and data analysis, as well as the original data used is included in the Author Contributions Checklist (ACC).
 }\fi

\section{Brain region details} \label{sec:appdata}
There are four primary brain regions, each containing smaller
subregions.
\begin{itemize}
\item \textbf{PFC-MSC}:
The prefrontal cortex (PFC) and primary motor-somatosensory
cortex (MSC) consist of six smaller regions:
primary motor cortex, primary
somatosensory cortex, ventral prefrontal cortex, medial prefrontal
cortex, dorsal prefrontal cortex and orbital prefrontal cortex.

\item \textbf{V1C, ITC, IPC, A1C, STC}: A region consisting of the primary
visual cortex (V1C), inferior temporal cortex (ITC), primary auditory cortex
(A1C), and superior temporal cortex (STC).

\item \textbf{STR, HIP, AMY}: A region consisting of the stratum (STR),
hippocampal anlage or hippocampus (HIP) and amygdala (AMY).

\item \textbf{MD, CBC}: A region consisting of the mediodorsal nucleus of the
thalamus (MD) and the cerebellar cortex (CD).
\end{itemize}

\section{Extension to the Stepdown method} \label{sec:accelerated}
One of the largest drawbacks of the Stepdown method  lies in its intensive computational
cost. For $r$ partitions, at most ${r \choose 2}$ bootstrap statistics need to be computed in each bootstrap trial,
each requiring a computational cost of $O(d^2\cdot n_p)$. 
In this section, we develop a computational extension to the Stepdown method that yields a more computationally
efficient algorithm as long as the test statistic $\hat{T}$ satisfies the \emph{triangle inequality} and the
number of variables $d$ is large.
That is, for any
bootstrap trial $b$ and for any partitions $i$, $j$ and $k$, we require that the bootstrap statistics satisfy
\begin{equation} \label{eq:triangle}
\hat{T}^{(b)}_{(i,k)} \leq \hat{T}^{(b)}_{(i,j)} + \hat{T}^{(b)}_{(j,k)}.
\end{equation}
This property can potentially save expensive calculations when calculating \eqref{eq:max_bootstrap} in \nameref{label:alg2}
by reducing the number of bootstrap statistics we need to explicitly calculate. Since we only care about the maximum bootstrap statistic $\hat{T}^{(b)}$ in each trial, the triangle inequality gives an upper bound on the bootstrap statistic $\hat{T}^{(b)}_{(i,k)}$ 
between partitions $i$ and $k$, leveraging bootstrap statistics already calculated within a specific bootstrap trial.
As we sequentially iterate through all pairs of partitions $(i,k)$,
if the upper bound for $\hat{T}^{(b)}_{(i,k)}$ is smaller than the current maximum bootstrap statistic within a specific bootstrap trial $b$,  we do not need
to explicitly compute $\hat{T}^{(b)}_{(i,k)}$. 

Unfortunately, the test statistic  \eqref{eq:test_stat} described in \Fref{ss:specification}
originally from  \cite{chang2015comparing} does not satisfy the triangle inequality \eqref{eq:triangle}.
Hence, we consider a new test statistic defined as
\begin{equation} \label{eq:test_stat_alternative}
\hat{T} = \max_{ij}\big(\hat{t}_{ij}\big) \quad \text{where
}\hat{t}_{ij}= \big|\hat{\sigma}_{X,ij}-\hat{\sigma}_{Y,ij}\big|, \quad i,j \in 1,\ldots, d,
\end{equation}
and we make a similar modification for its bootstrap counterpart, $\hat{T}^{(b)}$. 
It can easily be shown that the above bootstrap statistics satisfies the desired triangle inequality.
Additionally, using the techniques in \cite{chernozhukov2013gaussian}, it can be proven that this test statistic will still yield a
hypothesis test with asymptotic $1-\alpha$ coverage under the null, analogous to \eqref{eq:stepdown}.
We will call the Stepdown procedure that uses \eqref{eq:test_stat_alternative} the ``Accelerated Stepdown'' procedure.

To formalize how to take advantage of this triangle inequality property,
we describe a subroutine that leverages this 
property to compute $\hat{T}^{(b)}$ in \eqref{eq:max_bootstrap} by representing the individual bootstrap statistics 
$\hat{T}^{(b)}_{(i,j)}$ as weighted edges in a graph. The algorithm uses Dijsktra's algorithm to find the shortest path between vertices. This implicitly computes the upper bound in the bootstrap statistic between two partitions using the triangle inequality. 
This algorithm can provide substantial improvement in computational speed by leveraging the fact that determining the shortest path on a fully-dense graph has a computational complexity of $O(r^2)$, whereas computing $T^{(b)}_{(i,j)}$ has a computational complexity of
$O(d^2 \cdot n_p)$. 

\vspace{1em}
\textbf{\labelText{Algorithm 3}{label:alg3}: Distance metric-based procedure to compute
$\hat{T}^{(b)}$}
\begin{enumerate}
\item Form graph $G = (V,E)$ with $r$ nodes and all ${r \choose 2}$ edges, and initialize each edge to have a 
weight of infinity. 
\item Arbitrarily construct a spanning tree $\mathcal{T}$ and compute all $\hat{T}^{(b)}_{(i,j)}$'s corresponding to edges $(i,j) \in \mathcal{T}$. Record $z = \max_{(i,j) \in \mathcal{T}} \hat{T}^{(b)}_{(i,j)}$.
\item Construct a set of edges $\mathcal{S} = \mathcal{L} \backslash \mathcal{T}$ which represents
the bootstrap statistics between specific pairs of partitions that have yet to be computed.
\item While $\mathcal{S}$ is not empty:
\begin{enumerate}
\item Arbitrarily select an edge $(i,j) \in \mathcal{S}$ and remove it from $\mathcal{S}$. Compute the shortest-path distance from vertex $i$ to $j$ in $G$. 
\item If the shortest-path distance is larger than $z$, update the edge $(i,j)$ to have weight $\hat{T}^{(b)}_{(i,j)}$, and update $z$ to be $\max(z, \hat{T}^{(b)}_{(i,j)})$.
\end{enumerate}
\item Return $z$.
\end{enumerate}

As we will see in \Fref{sec:additional_sim}, while \eqref{eq:test_stat_alternative} can take advantage of this computational speedup, it yields a much less powerful test when compared to test using \eqref{eq:test_stat}.
This is intuitive, as \eqref{eq:test_stat_alternative} does not normalize by the
sum of the empirical variances, unlike \eqref{eq:test_stat}.
Hence, we do not use the Accelerated Stepdown procedure within COBS when analyzing the BrainSpan dataset
in this paper. However, we believe there are potentially other settings outside of covariance testing
where this computational speedup idea can be utilized more effectively.
We leave this as direction for future work.

\section{Details of algorithms to find quasi-cliques} \label{sec:quasiclique}

The first subsection remarks on possible extensions to the  clique-based selection method described in \nameref{label:alg4}.
The second subsection describes the three other algorithm used in \Fref{sec:simulation} for us to compare against.
Throughout this section, for a generic graph $G$, we use $V$ to denote the
set of vertices in $G$, $G_S$ to denote a subgraph formed by a vertex set $S \subseteq V$,
and $E(G)$ to denote the number of edges in $G$.

\subsection{Extensions to clique-based selection method} \label{ss:quasiclique}

We mention two extensions to clique-based selection method (\nameref{label:alg4}) that can be useful in practice.
\begin{itemize}
\item \textbf{Initializing algorithm around a desired set of vertices}: 
In certain cases, the user would want the $\gamma$-quasi-clique to be initialized around a desired subset of vertices
in $G = (V,E)$. For instance, in our setting, since \cite{liu2014network} applies DAWN to the 10 partitions in Window 1B,
it is natural for us to encourage COBS to select as many partitions in Window 1B as possible to enable a meaningful
comparison. 

To resolve this, first, we run \nameref{label:alg4} at the desired level $\gamma$ on $G_S$. 
This would output a subset of vertices $S_{\text{core}} \subseteq S$ that form the largest $\gamma$-quasi-clique in $G_S$. 
Then, we run  \nameref{label:alg4} at the same level $\gamma$ on the full graph $G$ but perform an additional operation after (2.): after $\mathcal{Q}$
is initialized with all maximal cliques in $G$, we check each vertex set $A \in \mathcal{Q}$ if $A \cup S_{\text{core}}$
forms a $\gamma$-quasi-clique. If yes, we replace $A$ with $A \cup S_{\text{core}}$ in $\mathcal{Q}$. If not, we remove $A$ from $\mathcal{Q}$. The algorithm then proceeds to (3.) as usual.
By applying this simple change, we are ensured the returned vertex set by \nameref{label:alg4} contains $S_{\text{core}}$.

\item \textbf{Post-processing the returned vertex set}: In certain cases, the returned vertex set of \nameref{label:alg4} has 
a few vertices with a very low degree when compared to the other vertices. 
To resolve this, we post-process this vertex set by removing vertices that are connected to less than half 
the other vertices in the returned set. 
\end{itemize}

In this paper, we use the initialization extension only when analyzing the BrainSpan dataset in \Fref{sec:brainspan},
where we initialize the largest quasi-clique around the 10 partitions in Window 1B.

\subsection{Overview of other algorithms}

We overview the three algorithms introduced in \Fref{sec:simulation}
that  are designed to find large quasi-cliques.

\begin{itemize}

\item \cite{chen2010dense}: This algorithm recursively splits a graph $G$ into two 
in a hierarchical-clustering type approach with respect to a carefully constructed weight matrix. 
This forms a tree-type data structure, and then the algorithm scans the tree in a breath-first-search type fashion for
the largest subgraph with an edge density larger than $\gamma$. 

\item \cite{tsourakakis2013denser}: This algorithm
performs a local search by adding vertices mypoically
and then removing vertices occasionally until no more myopic improvements can be made. 
Specifically, it first initializes the set $S$ of vertices to contain a vertex that maximizes the
ratio between the number of triangles and the degree,
and includes all of the neighbors of said vertex. Then algorithm iteratively tries to incrementally
improve the $f_{\gamma}(S) = E(G_S) - \gamma {|S| \choose 2}$ as much as possible by
adding neighbors of $S$. When it is no longer able to improve $f_{\gamma}(S)$,
the algorithm tries removing a vertex from $S$ to improve $f_{\gamma}(S)$. The algorithm then iterates
between such adding and removing vertices from $S$ for a fixed number of iterations.

\item Spectral clustering: While many different community detection methods for random graphs now exist (for example,
see \cite{abbe2017community} and \cite{athreya2017statistical} and the references within), we choose 
spectral clustering as described in \cite{lei2015consistency} as a prototypical example of
how many of such methods fail to demonstrate the monotone property as described in \Fref{ss:selection}.
Specifically, this method applies K-means clustering to the top $K$ eigenvectors of the adjacency matrix,
where $K$ is a tuning parameter to specify. 
To find large quasi-cliques, we iteratively try spectral clustering for a range of $K$'s (i.e., $K = 2,\ldots, 5$),
and for each detected cluster in any of the estimated clusterings, 
we compute if the corresponding vertices of said cluster forms a $\gamma$-quasi-clique. 
If any $\gamma$-quasi-clique is found, we return the largest  $\gamma$-quasi-clique discovered in this fashion.

\end{itemize}

\section{Formal description of simulation setup} \label{sec:simulation_app}
We say a multivariate vector $\bX \in \mathbb{R}^d$ is distributed based a nonparanormal distribution
with proxy mean vector $\bmu$, proxy covariance matrix $\bSigma$, and monotonic and differentiable functions $f_1,\ldots,f_d$ if the density of $\bX$ is
\begin{equation} \label{eq:nonparanormal}
p(\bx) = \frac{1}{(2\pi)^{d/2}|\bSigma|^{1/2}}\exp\Big\{
-\frac{1}{2}\big(f(\bx) -\bmu\big)^\top \bSigma \big(f(\bx) - \bmu\big)
\Big\}\prod_{j=1}^{d}\big|f'_j(x_j)\big|,
\end{equation}
where $f(\bx) = (f_1(x_1), \ldots, f_d(x_d))$.
This is defined in \cite{liu2009nonparanormal}.
 In our simulation suite, we set $\bmu = 0$. 
 Let this distribution be denoted as $\text{NPN}(\mathbf{0}, \bSigma, f)$.
In the next two subsections, we formalize the details of $\bSigma$ and $f_1,\ldots,f_d$.

\subsection{Details on proxy covariance matrices $\mathbf{\Sigma}$}

The following three bullet points detail the construction of $\bSigma^{(1)}$, $\bSigma^{(2)}$ and $\bSigma^{(3)}$ respectively. As mentioned in \Fref{sec:simulation}, $\beta \in [0,1]$ is a user-defined parameter that controls the dissimilarity
among these three matrices.

\begin{itemize}
\item \textbf{Construction of $\bSigma^{(1)}$}: As mentioned in \Fref{sec:simulation}, $\bSigma^{(1)} \in \mathbb{R}_+^{d\times d}$ follows an SBM with two equally-sized clusters. Specifically, the first cluster contains indices $1, \ldots, \lfloor d/2 \rfloor$ and the 
second cluster contains indices $\lfloor d/2 \rfloor + 1,\ldots, d$. Then, we construct $\bSigma^{(1)}$ where
\begin{equation} \label{eq:sbm}
\Sigma^{(1)}_{ij} = 
\begin{cases}
1 &\quad \text{if }i=j,\\
a &\quad \text{if }i \neq j, \;i \text{ is in the same cluster as }j,\\
b  &\quad \text{if }i \neq j, \;i \text{ is not in the same cluster as }j,
\end{cases}
\end{equation}
for all $i,j \in 1,\ldots, d$ 
and $a = 0.9$ and $b = 0.1$.

\item \textbf{Construction of $\bSigma^{(2)}$}: $\bSigma^{(2)}$ is constructed the same as $\bSigma^{(1)}$, except
\[
a = 0.9 - \beta \cdot 0.4, \quad \text{and} \quad b = 0.1 + \beta \cdot 0.4.
\]
When $\beta = 1$, this means that $\bSigma^{(2)}$ is a matrix with 0.5 everywhere along the off-diagonal.

\item \textbf{Construction of $\bSigma^{(3)}$}: $\bSigma^{(3)}$ is constructed in a similar way to $\bSigma^{(1)}$, except there are
three clusters. The first cluster contains indices $1, \ldots, \lfloor \beta \cdot d/6 \rfloor, \lfloor d/2 \rfloor + 1, \ldots,
 \lfloor d/2 +\beta \cdot d/6 \rfloor$.
 The second cluster contains indices $\lfloor \beta \cdot d/6 \rfloor + 1, \ldots, \lfloor d/2 \rfloor$.
 The third cluster contains indices $ \lfloor d/2 +\beta \cdot d/6 \rfloor + 1, \ldots, d$. 
 Observe that this partitions $1,\ldots,d$, and when $\beta=1$, this results in three clusters of the roughly the same size.
 We then construct $\bSigma^{(3)}$ using \eqref{eq:sbm} but using these three clusters.

\end{itemize}

\subsection{Details on functions $f_1, \ldots, f_d$}
At a high-level, these functions $f_1,\ldots,f_d$
ensure that these marginal distributions of our sampled nonparanormal random
variables
are similar to the marginal distributions of the BrainSpan data.
These marginal distributions are constructed in the following way.
We first randomly sample $d$ variables (i.e., genes) uniformly from the BrainSpan dataset, $\{g_1,\ldots,g_d\}\subseteq \{1,\ldots,n\}$. Next, for each $j$, let $\hat{p}_{g_j}$ denote the kernel density estimate of variable $g_j$
in the BrainSpan dataset, using the default bandwidth selection used by the \texttt{stats::density} function in \texttt{R}.

We now formalize how to construct $f_1, \ldots, f_d$. As described in \cite{liu2009nonparanormal},
we actually construct the inverse of these functions
$f^{-1}_1, \ldots, f^{-1}_d$ as they are more amendable for sampling, which must exist since $f_1,\ldots,f_d$ are monotonic and differentiable.
Recall that $\bmu = \mathbf{0}$.
We first sample a vector $\bz = (z_1,\ldots,z_d)$ from a Gaussian distribution $N(\mathbf{0}, \bSigma)$.
Let $\Phi(t; P)$ denote the cumulative distribution function evaluated at $t$ for a univariate density $P$.
For any $j \in 1,\ldots,d$, we construct $f^{-1}_j$ such that
\[
\Phi\big(t; N(0, \Sigma_{jj})\big) = \Phi\big( f^{-1}_j(t); \hat{p}_{g_j}\big), \quad \forall t \in \mathbb{R}.
\]
That is, we construct $f^{-1}_j$ so that $z_j$ is at the same quantile with respect to $N(0, \Sigma_{jj})$ as
$f^{-1}_j(z_j)$ is with respect to the kernel density estimate $\hat{p}_{g_j}$. 
Notice that by constructing $f^{-1}_1,\ldots, f^{-1}_d$ in this fashion, each function is monotone and differentiable.
We then set
\[
\bx = \big(x_1,\ldots,x_d\big) = \big(
f^{-1}_1(z_1),\ldots, f^{-1}_d(z_d)
\big)
\]
as one sample from the nonparanormal distribution $\text{NPN}(\mathbf{0}, \bSigma, f)$.

Notice that by introducing non-Gaussianity into our simulation suite in this fashion, we ensure that the marginal distribution of all 
$r$ partitions resemble the BrainSpan dataset, and also ensure that the first $r_1$ partitions still are drawn from the same population covariance matrix.
Also, by generating data in this fashion, we are able to obtain complicated dependencies between the mean and
variance, as well as observe multi-modal distributions and heavier-tailed distributions compared to the Gaussian.
See  \cite{liu2009nonparanormal} for a more detailed discussion.

\subsection{Example of sampled nonparanormal distribution}
\begin{figure}[H]
  \centering
  \includegraphics[width=400px]{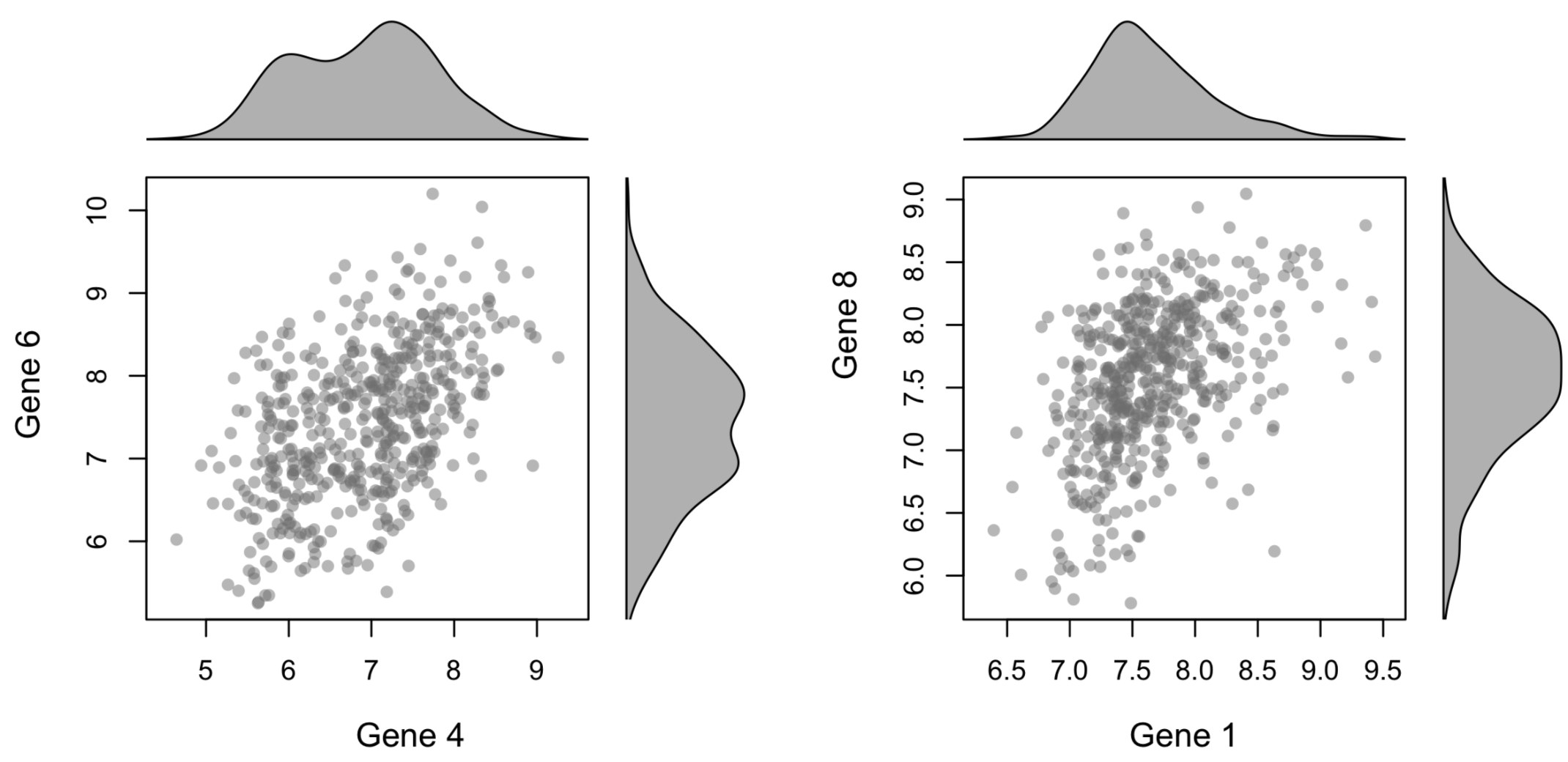}
    \caption
   {\small Two scatter plots of bivariate distributions sampled from the nonparanormal for $\beta=0$.
   The densities shown on the top and the right of each plot represents the targeted kernel density
   estimates from the BrainSpan data that the nonparanormal is sampling from, captured by $f_1, \ldots, f_d$.}
    \label{fig:nonparanormal_example}
\end{figure}

We provide a visual illustration of what the sampled nonparanormal distribution could look like.
We sample 375 samples from $\text{NPN}(\mathbf{0}, \bSigma^{(1)}, f)$ when $\beta = 0$, and plot two
of the resulting pairwise scatterplots in \Fref{fig:nonparanormal_example}.
We can think of the 375 samples as equivalent to 
aggregating all $r=25$ partitions together,  each having $n=15$ samples.
 These two scatterplots show 
that the nonparanormal can display multiple modes marginally or heavier tails.

\section{Additional simulation results} \label{sec:additional_sim}

\subsection{Covariance homogeneity diagnostic in simulation}

In this section, we apply the diagnostic developed in \Fref{sec:elementary}
to the simulation suite described in \Fref{sec:simulation}. 
Our goal is to determine how the QQ-plots behave as the selected partitions become less
homogeneous. As done in \Fref{sec:simulation}, we consider four partition selection
strategies: COBS (using $\alpha = 0.1$ and $\gamma = 0.95$), Base (which selects
 3 partitions contain samples drawn from 
the nonparanormal distribution with proxy covariance $\bSigma^{(1)}$, 
while the other 2 partitions contain samples from each of the remaining two distributions),
All (which selects all $r$ partitions) and Oracle  (which selects 
exactly the $r_1$ partitions containing samples drawn the nonparanormal distribution
 with proxy covariance $\bSigma^{(1)}$).
 
\begin{figure}[H]
  \centering
  \includegraphics[width=450px]{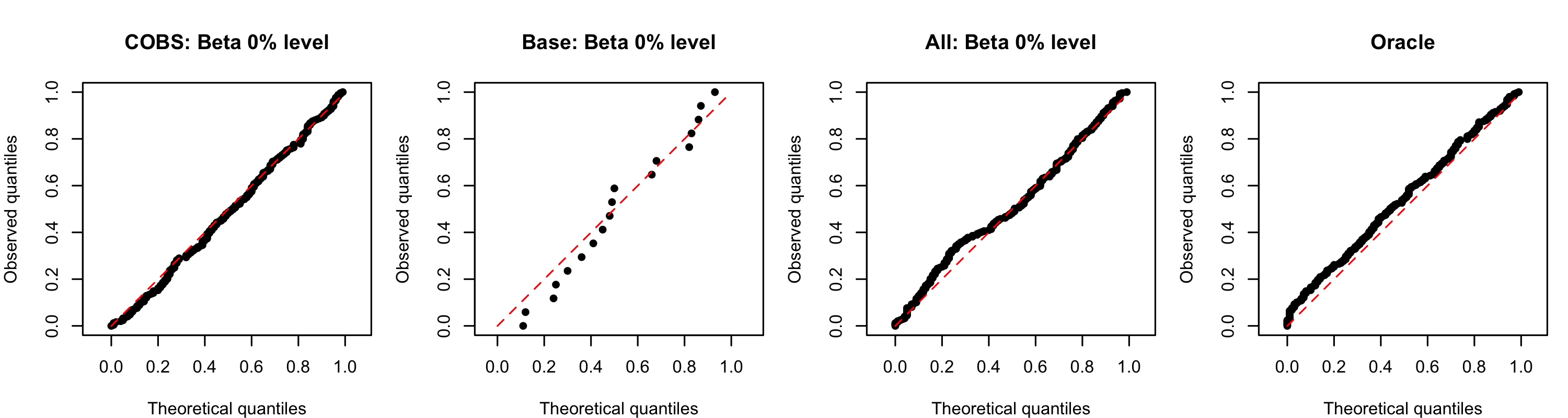}
    \includegraphics[width=450px]{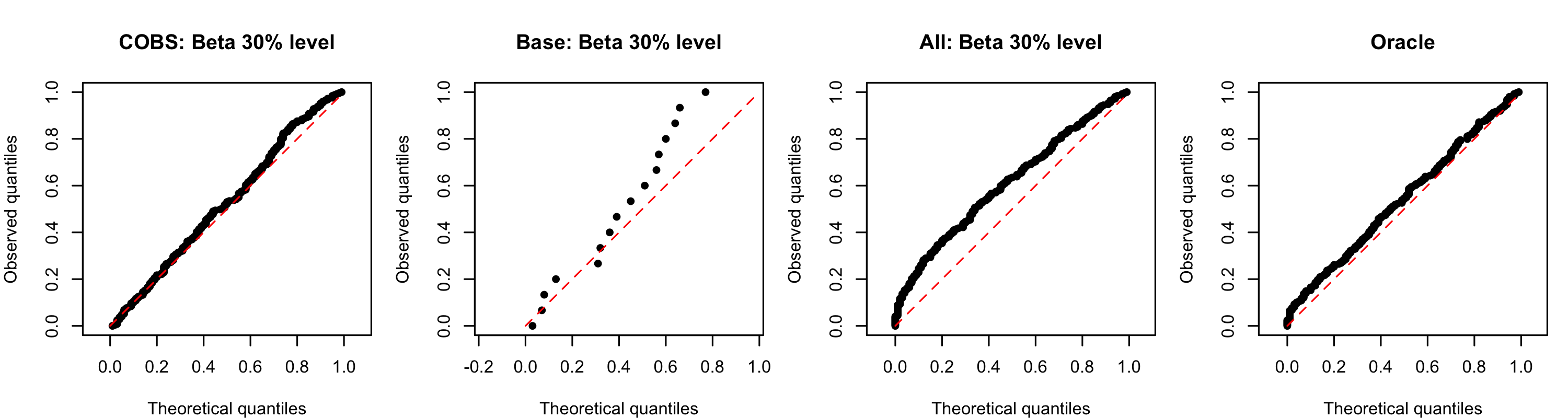}
  \caption
   {\small QQ-plots from the covariance homogeneity diagnostic using four different selection procedures:
   COBS (left-most), Base (center left), All (center right) and Oracle (right-most). 
   The top row represents the simulation setting where $\beta = 0$, while the second row represents the
   simulation setting where $\beta = 0.3$.
   The plots are created from one instance of COBS, Base, All and Oracle procedures, and 250 trials are used within the covariance homogeneity diagnostic.}
    \label{fig:app_goodness_1}
\end{figure}

\begin{figure}[H]
  \centering
  \includegraphics[width=450px]{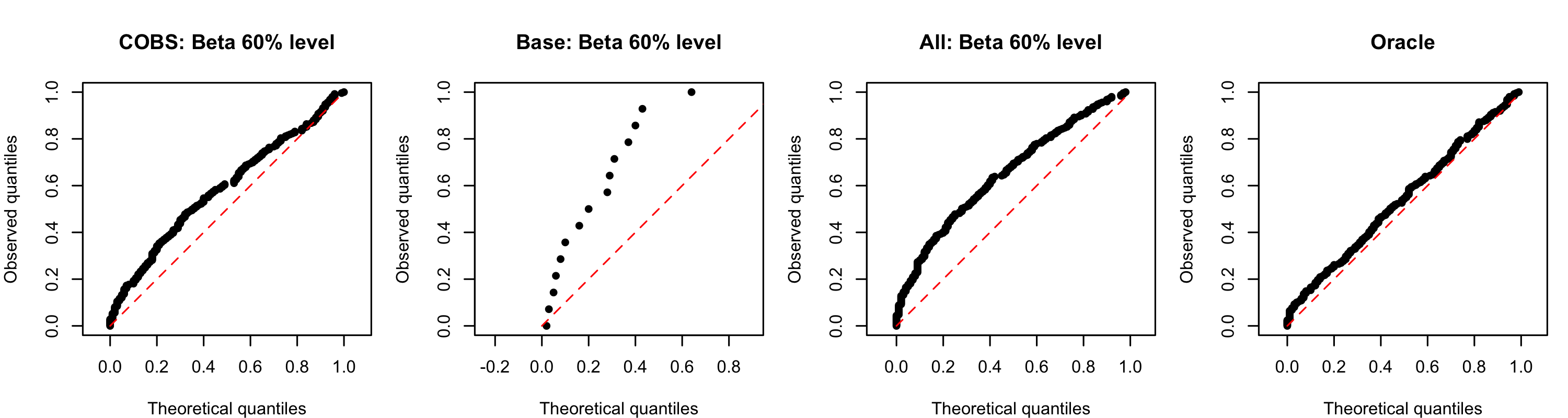}
    \includegraphics[width=450px]{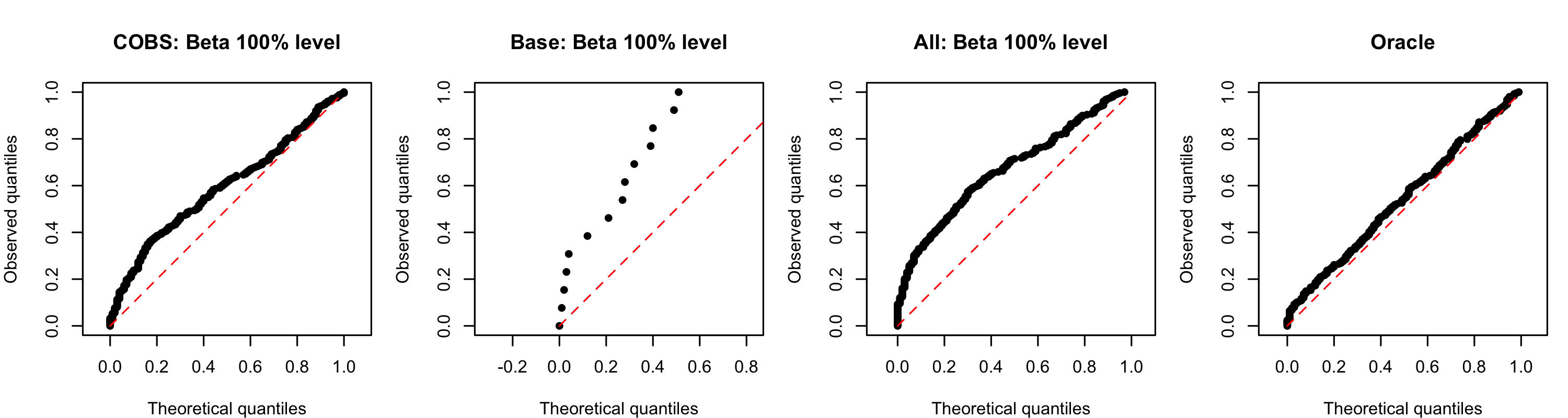}  
    \caption
   {\small QQ-plots derived in a similar way as in \Fref{fig:app_goodness_2}. However, in this plot, 
   the top row represents the simulation setting where $\beta = 0.6$, while the second row represents the
   simulation setting where $\beta = 1$.}
    \label{fig:app_goodness_2}
\end{figure}

 We see in \Fref{fig:app_goodness_1} and \Fref{fig:app_goodness_2} that the QQ-plot is a reasonable
 diagnostic in this simulation suite.
 Between these two figures, we vary $\beta$ among 0, 0.3, 0.6 and 1. 
 We notice that as $\beta$ increases, the QQ-plot derived from COBS remains relative uniform, similar to that of the Oracle. When $\beta = 1$, COBS selects one erroneous partition in this particular trial shown, which results in the
 QQ-plot showing a deviation away from uniform.
 The QQ-plots derived from the Base procedure looks relative uniform when $\beta = 0$ (which is to be expected,
 as all $r$ partitions share the same covariance matrix when $\beta = 0$), but quickly has QQ-plots that deviate
 from uniform as $\beta$ increases.
 Note that the since the Base procedure selects only 5 partitions, there are a limited number of ways to
 split the partitions into two groups, which yields a limited number of points in the QQ-plot.
 The QQ-plots derived from the All procedure follow a similar trend as the Base procedure, but not as
 severe. These plots match the findings shown in \Fref{fig:snr}.

\subsection{Simulation under Gaussian setting}
While the simulations in \Fref{sec:simulation} use nonparanormal distributions,
we demonstrate that similar results hold for Gaussian distributions. This demonstrates
that there is nothing particularly special about the nonparanormal or the Gaussian distribution
that enable COBS to work well, and suggests COBS can work in much more general settings.
Specifically, in this simuation suite, everything is the same as in  \Fref{sec:simulation}, except all the
functions $f_1,\ldots,f_d$ are set to be the identity function. Hence, this means that the 
first $r_1$ partitions are drawn from Gaussian distributions with covariance $\bSigma^{(1)}$,
the next $r_2$ partitions are drawn from Gaussian distributions with covariance $\bSigma^{(2)}$,
and so on.

\begin{figure}[H]
  \centering
  \includegraphics[width=400px]{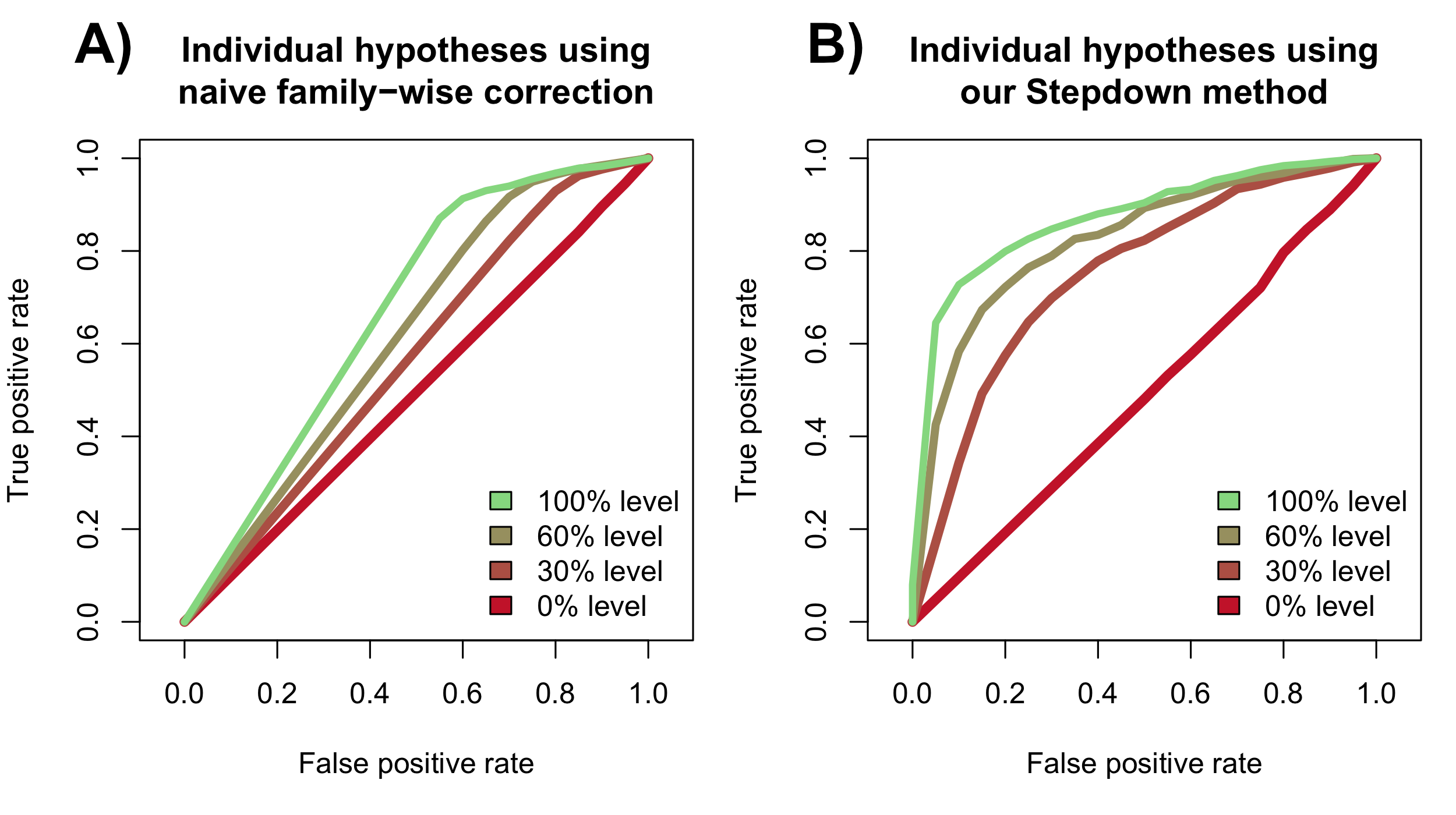}
    \vspace{-1em}
  \caption
   {\small ROC curves for the hypothesis under the Gaussian setting. These plots are similar to those in \Fref{fig:roc_hypothesis}.
   }
    \label{fig:app_gaussian_1}
\end{figure}

When we use Bonferroni or the Stepdown method in this Gaussian setting, we observe ROC curves
for the individual hypotheses that strongly resemble \Fref{fig:roc_hypothesis}. This is shown in \Fref{fig:app_gaussian_1}.

\begin{figure}[H]
  \centering
  \includegraphics[width=400px]{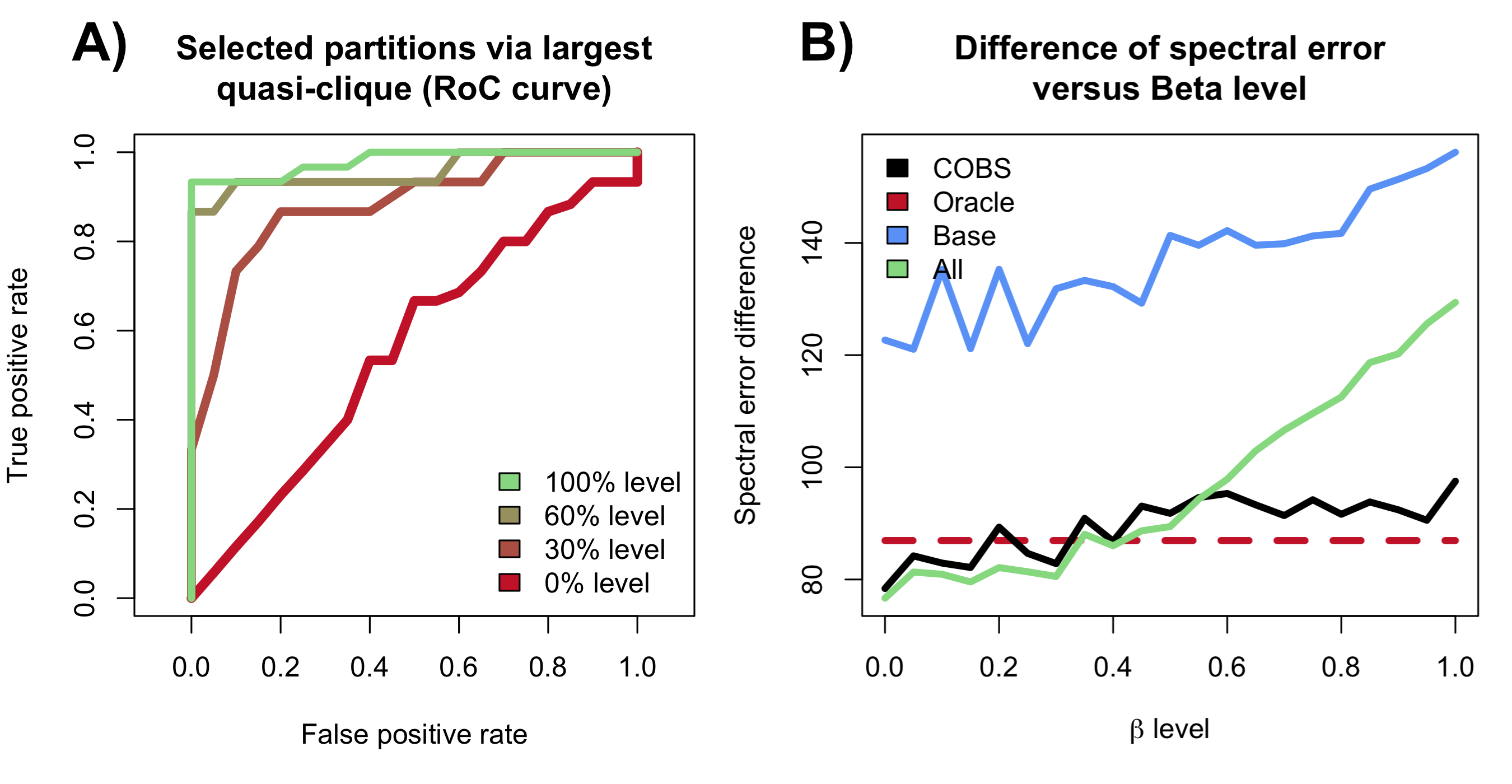}
    \vspace{-1em}
  \caption
   {\small A) ROC curves for the partitions selected by COBS under the Gaussian setting. This plot is similar to 
   \Fref{fig:snr}A.   
   B) Mean spectral error of the estimated covariance matrix for varying $\beta$ level under the Gaussian setting.
   This plot is similar to 
   \Fref{fig:snr}B.
   }
    \label{fig:app_gaussian_2}
\end{figure}

Similarly, when we use COBS to select partitions, the ROC curves as well as the spectral error curves
strongly resemble \Fref{fig:snr}A and B. This is shown in \Fref{fig:app_gaussian_2}.

\subsection{Simulation using Accelerated Stepdown}
In this subsection, we apply the Accelerated Stepdown procedure described in \Fref{sec:accelerated}
within the COBS procedure in the simulation setting described in \Fref{sec:simulation}.
Specifically, we use the test statistic \eqref{eq:test_stat_alternative} and 
analogous bootstrap statistics, but keep all other parts of the simulation suite the same. 

\begin{figure}[H]
  \centering
  \includegraphics[width=400px]{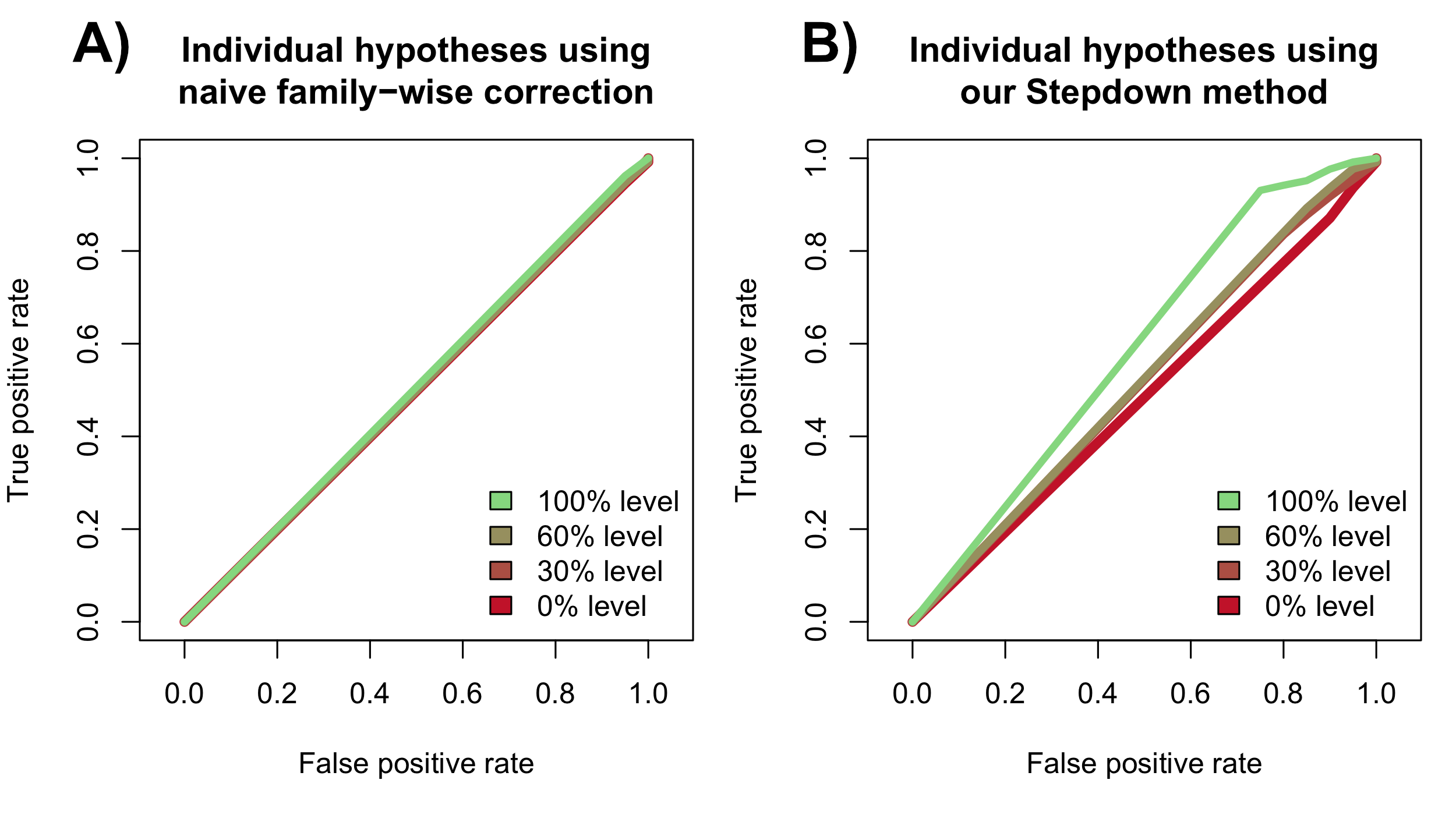}
    \vspace{-1em}
  \caption
   {\small ROC curves for the hypothesis using the Accelerated Stepdown procedure
   described in \Fref{sec:accelerated} in the nonparanormal setting. These plots are set up in the 
   same as in \Fref{fig:roc_hypothesis}.   }
    \label{fig:app_accelerated_1}
\end{figure}

When we plot the ROC curve for the individual hypotheses in \Fref{fig:app_accelerated_1}, we already notice a dramatic loss of power 
when compared to its original counterpart using the test statistic \eqref{eq:test_stat} shown in  \Fref{fig:roc_hypothesis}.
In fact, it seems like the Bonferroni procedure has almost no power at all, even when $\beta = 1$.

\begin{figure}[H]
  \centering
  \includegraphics[width=400px]{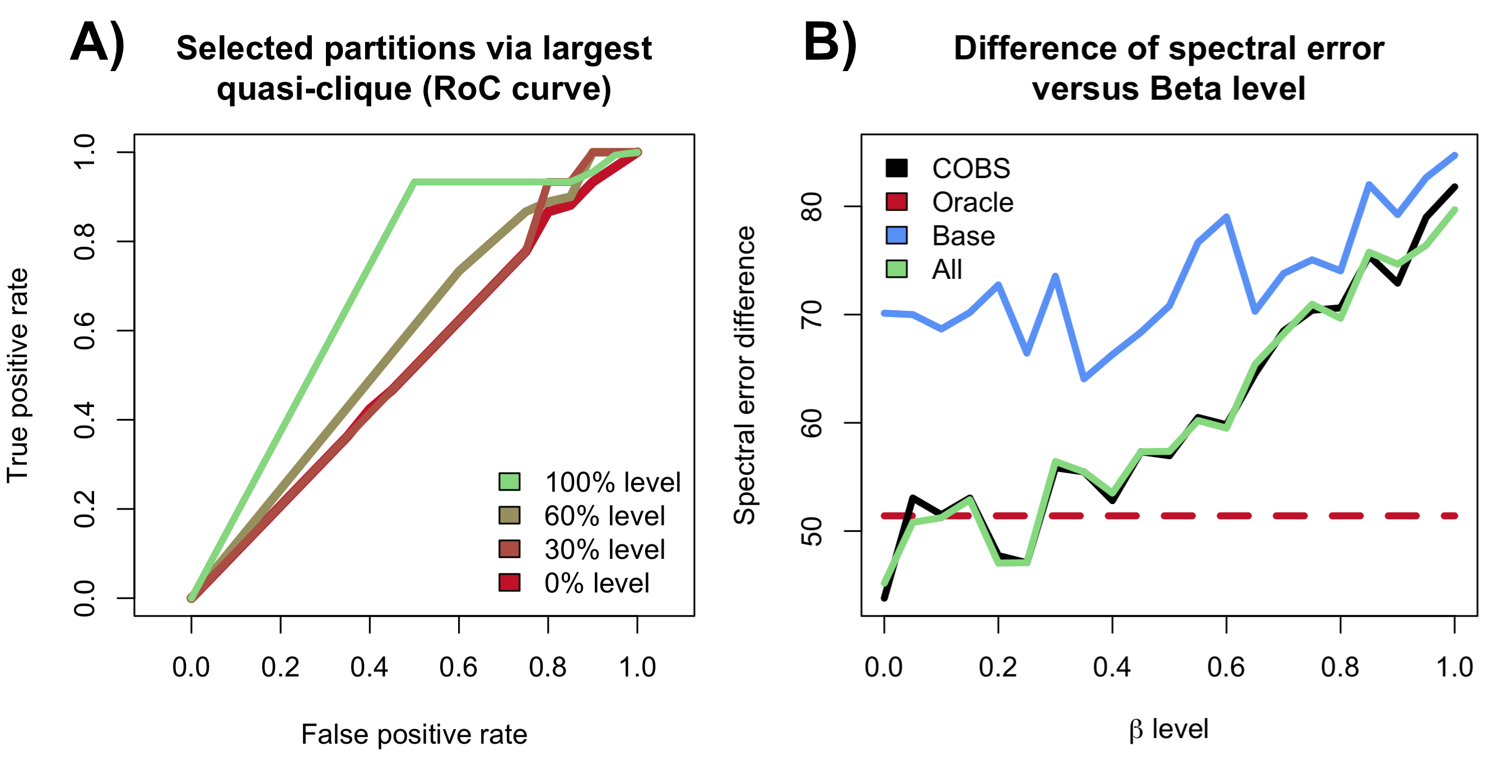}
    \vspace{-1em}
  \caption
   {\small A) ROC curves for the partitions selected by COBS using the Accelerated Stepdown procedure
   described in \Fref{sec:accelerated} in the nonparanormal setting. This plot is set up in the same way as in 
   \Fref{fig:snr}A.   
   B) Mean spectral error of the estimated covariance matrix for varying $\beta$ level using the Accelerated Stepdown procedure in the nonparanormal setting.
   This plot is set up in the same way as in    \Fref{fig:snr}B.
   }
    \label{fig:app_accelerated_2}
\end{figure}

Due to the loss of power for the individual hypotheses, we observe a loss of power for the selected partitions
as well (\Fref{fig:app_accelerated_2}A) and spectral errors that strongly resemble selecting all the partitions
(\Fref{fig:app_accelerated_2}B).

\section{Additional details on BrainSpan analysis} \label{sec:additional_analysis}

The first subsection describes the analysis pipeline we used throughout \Fref{sec:brainspan} in
more detail. The second subsection describes the two distance metrics used in \Fref{ss:val}.
The third subsection describes additional results alluded to in \Fref{ss:val}.

\subsection{Description of analysis pipeline}
We now summarize the pipeline used in \Fref{sec:brainspan} for clarity.

\begin{enumerate}

\item \textbf{Screening of genes}: This is the step described in \Fref{ss:screening}, derived from \cite{liu2014network}. 
We first select all
genes whose p-value in the older TADA dataset \citep{de2014synaptic} is less than 0.01.
Then, we rank all remaining genes by their maximum Pearson correlation in magnitude with any of the
formerly selected genes in decreasing order based on the BrainSpan partitions within Window 1B aggregated.
We select genes based on this ranking in order until we have selected a combined total of $d=3500$ genes.
We analyze all 125 partitions using only these $d$ genes for the remainder of the analysis.

\item \textbf{Applying COBS}: This is the two-staged procedure we developed in this paper, detailed in \Fref{sec:method}.
In the first stage, we apply the Stepdown procedure using $\alpha = 0.1$. 
In the second stage, we select the clique-based selection method where $\gamma =0.95$, as well as using 
both extensions discussed in \Fref{ss:quasiclique}. 
This results in 24 selected partitions within the BrainSpan dataset, as detailed in \Fref{ss:partition}.
We then combine all the 24 selected partitions to form a dataset $\mathbb{X} \in \mathbb{R}^{n \times d}$ with $n=272$ microarray samples
and $d$ genes to be used for the remainder of the analysis.

\item \textbf{Estimating the Gaussian graphical model}: 
This step is described in \Fref{ss:gene_network} and is the same as in \cite{liu2014network}. We fit a 
Gaussian graphical model using neighborhood selection \citep{meinshausen2006high} based on $\mathbb{X}$, where the tuning
parameter $\lambda$ (which controls the sparsity of the graphical model) 
is chosen such that the resulting graph has high scale-free index
as well as a comparable number of edges to the estimated graph when COBS is
not used. 
This choice of $\lambda$ is detailed at the end of this subsection. 
We defer the remaining estimation and computation details to \cite{liu2014network}.
We denote the adjacency matrix 
of the estimated graphical model as $\hat{\bA} \in \{0,1\}^{d \times d}$.

\item \textbf{Estimating the HMRF}:
This step is also described in \Fref{ss:gene_network} and is the same as in \cite{liu2014network}. 
We briefly summarize this step here, as it is less common in the statistical literature.
Let $\bZ \in \mathbb{R}^d$ denote the Z-scores for the selected
genes, derived from the TADA scores in \cite{de2014synaptic}. 
We model $\bZ$ using a HMRF, where for each gene in $j = 1,\ldots,d$, $Z_j$ is
an i.i.d. random variable drawn from a mixture of  two Gaussians,
\[
Z_j \sim \mathbb{P}(I_j = 0)N(0, \sigma^2) + \mathbb{P}(I_j = 1)N(\mu, \sigma^2),
\]
where $I_j \in \{0,1\}$ is an unobserved Bernoulli random variable and $\mu \in \mathbb{R}$ and $\sigma \in \mathbb{R}_+$ are two unknown scalars to be estimated. The first Gaussian distribution represents
the Z-scores for genes that are not associated with ASD, and the second Gaussian distribution
represents the Z-scores for risk genes.
The distribution of entire vector $\bI \in \{0,1\}^d$ follows an Ising model with probability mass function,
\[
\mathbb{P}(\bI = \beeta) \propto \exp\Big(b \cdot \sum_{j=1}^{d} \eta_j + c \cdot \beeta^T\hat{\bA} \beeta\Big),
\]
for any $\beeta \in \{0,1\}^d$ and  two unknown scalars $b,c\in \mathbb{R}$ to be estimated. An EM algorithm is used
to fit this HMRF model, and we obtain the 
estimated posterior probability $\hat{p}_j = \mathbb{P}(I_j = 0 | \bZ)$, representing
the probability gene $j$ is not a risk gene given the risk scores.
We defer the estimation and computation details to \cite{liu2014network}.

\item \textbf{Applying Bayesian FDR}: This step is also described in \Fref{ss:gene_network} and is the same as in \cite{liu2014network}. We apply a procedure \citep{muller2006fdr}  to $\hat{\bp}$ to select a set of genes
where the Bayesian FDR is controlled at level $10\%$.
We defer the computation details to \cite{liu2014network}.
This results in the set of 209 detected risk genes detailed in \Fref{ss:val}.
\end{enumerate}

\paragraph{Usage of  \cite{de2014synaptic}.} We note that the older risk scores dataset  \citep{de2014synaptic} is used twice, once in the screening stage (Step 1 above) and
again to estimate the parameters of the HMRF (Step 4 above).
As argued by \cite{liu2014network}, it is important for this dataset to be the same in both steps,
as the goal of DAWN is to boost the power of the risk scores by a ``guilt-by-association'' 
strategy. 
Hence, it is important to ensure the genes with low TADA scores remain in the analysis after screening, so they
can implicate genes with TADA scores that are not as low.

\paragraph{Choice of $\lambda$.} We use the following procedure to tune $\lambda$ when estimating the
Gaussian graphical model using only the 10 partitions from Window 1B as well as when using the 24 partitions
selected by COBS.
We tune $\lambda$ on a grid between 0.05 and 0.1, equally spaced into 15 values, for both graphical models.
Our criteria for selecting $\lambda$ within this grid is inspired by
\cite{liu2014network}, who use a \emph{scale-free index}, a number between 0 and 1 that measures how well
the graph follows a power law.
Specifically, we ensure the scale-free indices from both graphical models are approximately comparable as well as that
both estimated graphical models have about 10,000 edges. Our focus on this number of edges comes from \cite{liu2014network}, which estimated a graphical
model with 10,065 edges.
By ensuring both of our estimated graphical models have around 10,000 edges, we are able to ensure that
both graphical models pass roughly the same amount of information into the HMRF stage of DAWN.

Using this procedure, we set $\lambda = 0.05$ when estimating the graphical model 
using only the 10 partitions from Window 1B (for 9990 edges and a scale-free index of 0.77)
and $\lambda = 0.064$ when estimating the graphical model using the 24 partitions selected by COBS (for 9142 edges
and scale-free index of 0.83).

\subsection{Methods to measure distance of two nodes in a graph}

As alluded to in \Fref{ss:val}, the shortest path distance and the commute distance do not seem like 
appropriate candidates to measure the distance between two genes (i.e., vertices) in a  gene co-expression network 
(i.e., graph)
due to the fact that the network estimated in the Window 1B analysis has more edges than in the COBS analysis
(9990 and 9142 edges respectively). Hence, both of these distance metrics would naturally favor the denser 
graph.

To overcome this problem, we use two distance metrics that we believe enable a more fair comparison.

\begin{itemize}
\item \textbf{Minimal spanning tree (MST) distance}: This is a natural alternative to measure the distance between
two vertices. Given a graph $G=(V,E)$, we first find the MST $G_{\text{(MST)}} \subseteq G$, and then compute the path distance
between the two vertices in $G_{\text{(MST)}}$.

\item \textbf{Graph root embedding distance}: A more statistically motivated way to measure the distance
between two vertices is to first embed all vertices $V$ into a latent space. As shown in \cite{lei2018network},
the graph root embedding is a natural candidate to do this, as it can theoretically represent a wide range of
random graphs. This is essentially a more sophisticated spectral embedding.
We first represent the graph $G$ as an adjacency matrix $\bA$, and compute the top-$k$ eigenvectors (corresponding both the largest $k$ positive eigenvalues and largest $k$ negative eigenvalues in magnitude).
Each vertex is then represented as a latent vector of length $2k$. The distance between two vertices
is then defined as the Euclidean distance between their corresponding latent vectors.
We defer the remaining details to  \cite{lei2018network}.

It is important to use both positive and negative eigenvalues since a scree plot reveals there are almost
the same number of positive and negative eigenvalues for the adjacency matrices estimated in both the COBS and Window 1B analyses.
\end{itemize}

\subsection{Additional results about closeness of genes}

We provide more details that the 102 genes detected by the 
newer TADA scores \citep{satterstrom2019large} are roughly 10\%-30\% 
closer to the 33 genes detected
in the older TADA scores \citep{de2014synaptic} in the gene network estimated in the COBS analysis
than  in the Window 1B analysis. 
We call the 33 genes detected in  \cite{de2014synaptic} as the De Rubeis genes,
and the 102 genes detected in \cite{satterstrom2019large} that are not part of the former 33 genes
as the Satterstrom genes.

\begin{figure}[H]
  \centering
  \includegraphics[width=400px]{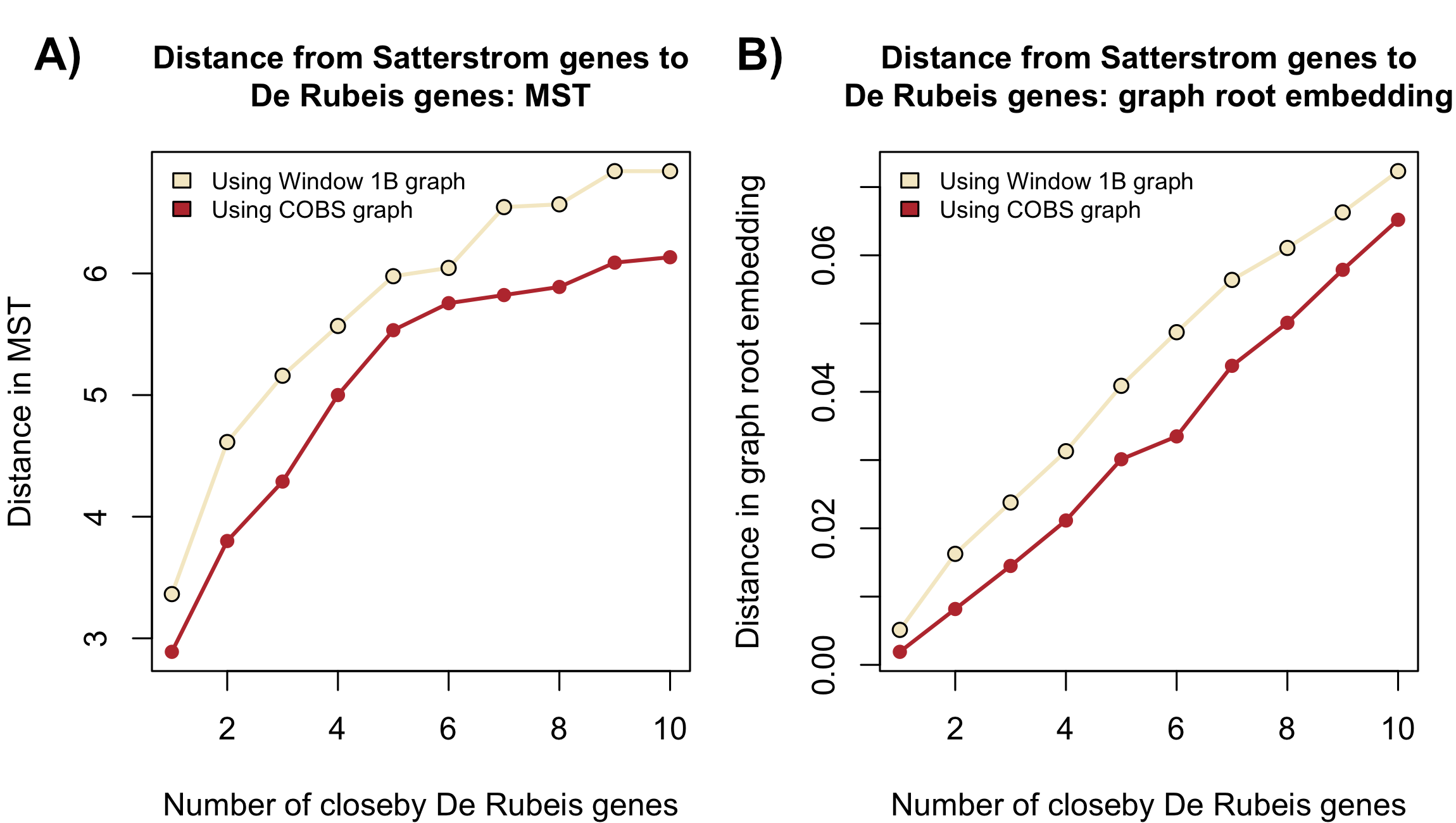}
  \caption
   {\small A) Average MST distance from a Satterstrom gene to the closest $k$ De Rubeis genes against $k$.
   B) Average graph root embedding distance from a Satterstrom gene to the closest De Rubeis genes against
   the half of the embedding dimension $k$.
   }
    \label{fig:app_distance}
\end{figure}

We use the MST distance defined above to ask: how far away are the closest $k$ De Rubeis genes from
any Satterstrom gene on average (mean). 
\Fref{fig:app_distance}A plots this average distance against $k$.
We use the graph root embedding distance to ask: how close is the nearest De Rubeis gene from
any Satterstrom gene on average (mean) when using an embedding of latent dimension $2k$.
\Fref{fig:app_distance}B plots this average distance against $k$. 
In both instances, regardless of how the parameter $k$ is chosen, the plot shows that
the Satterstrom genes are closer to the De Rubeis genes on average.
Both metrics show that the red curve is roughly 10\%-30\% lower than the pale curve across all values of $k$, hence
giving our stated result.

\end{document}